\documentclass[preprint,5p,times,twocolumn]{elsarticle}
\usepackage{epsfig}
\usepackage{color}
\usepackage{url}

\usepackage{algorithm}
\usepackage{algorithmic}
\usepackage{amsmath}
\usepackage{lmodern}
\usepackage{amssymb}
\usepackage{graphicx}
\usepackage{epstopdf}
\usepackage{color}
\usepackage{bm}

\journal{SPM 2019}

\definecolor{myRedColor}{RGB}{200, 0, 0}
\definecolor{myGreenColor}{RGB}{0, 125, 0}
\definecolor{myBlueColor}{RGB}{0, 0, 155}
\definecolor{myYellowColor}{RGB}{155,155, 0}
\definecolor{myCyanColor}{RGB}{0, 155, 155}
\definecolor{myMagentaColor}{RGB}{155, 0, 155}
\definecolor{myBrownColor}{RGB}{190, 100, 60}

\newcommand{\Reals}{\mbox{\normalsize{\sl I\mbox{\hspace{-0.03in}}R}}}

\newcommand{\Bspline}{B-spline}
\newcommand{\Bezier}{B\'{e}zier}
\newcommand{\NURB}{NURBs}

\newcommand{\microstruct}{\mbox{${\cal M}\hspace{-0.16in}{\cal M}$}}

\begin{document}

\begin{frontmatter}

\title{\bf Optimizing micro-tiles in micro-structures as a design paradigm}


\author{Pablo Antolin}
\address{\'{E}cole Polytechnique F\'{e}d\'{e}rale de Lausanne,
	Institute of Mathematics, Lausanne, Switzerland}
\author{Annalisa Buffa}
\address{\'{E}cole Polytechnique F\'{e}d\'{e}rale de Lausanne,
	 Institute of Mathematics, Lausanne, Switzerland and
	Istituto di Matematica Applicata e Tecnologie Informatiche
	'E. Magenes' (CNR), Pavia, Italy}
\author{Elaine Cohen}
\address{Department of Computer Science, University of Utah,
	SLC, Utah, USA}
\author{John F. Dannenhoffer}
\address{Mechanical \& Aerospace Engineering, Syracuse University,
	 Syracuse, NY, USA}
\author{Gershon Elber}
\address{Department of Computer Science
	Technion, Israel Institute of Technology,  Haifa 32000, Israel}
\author{Stefanie Elgeti}
\address{Chair for Computational Analysis of Technical Systems,
	 RWTH Aachen University, Aachen, Germany}
\author{Robert Haimes}
\address{Department of Aeronautics and Astronautics, MIT, Cambridge,
	Massachusetts, USA}
\author{Richard Riesenfeld}
\address{Department of Computer Science, University of Utah,
	SLC, Utah, USA}

\begin{abstract}

In recent years, new methods have been developed to synthesize complex
porous and micro-structured geometry in a variety of ways.  In this
work, we take these approaches one step further and present these
methods as an efficacious design paradigm. Specifically, complex
micro-structure geometry can be synthesized while optimizing certain
properties such as maximal heat exchange in heat exchangers, or
minimal weight under stress specifications.

By being able to adjust the geometry, the topology and/or the material
properties of individual tiles in the micro-structure, possibly in a
gradual way, a porous object can be synthesized that is
optimal with respect to the design specifications.  As part of this
work, we exemplify this paradigm on a variety of diverse applications.

\end{abstract}


\end{frontmatter}

%
%
%

\section{Introduction}
\label{sec-intro}

\begin{figure}
    \begin{center}
    \begin{tabular}{c}
	\mbox{\hspace{-0.15in}}
        \epsfig{file=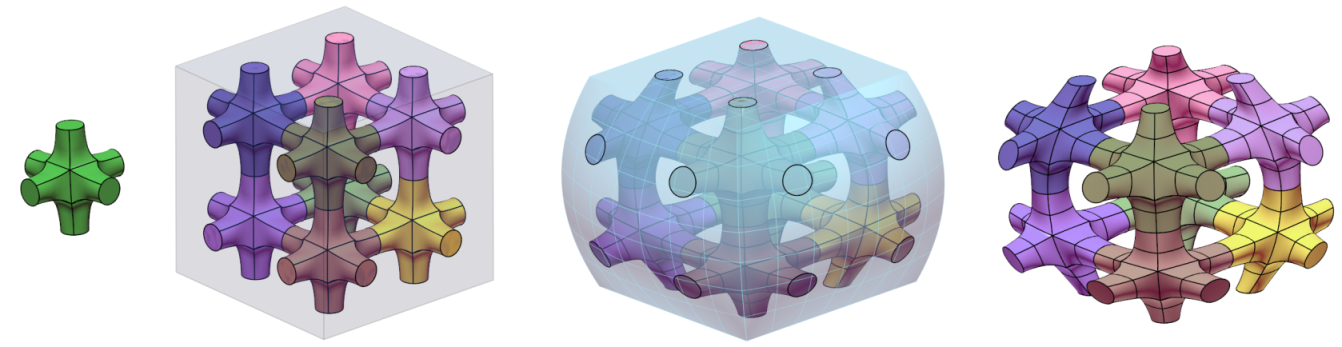,width=3.5in}
    \end{tabular}
    \begin{picture}(0,0)
        \put(-120, 15){(a)}
        \put(-120, 52){$M$}
        \put( -97,  5){(b)}
        \put( -42, 60){$D$}
        \put(  35,  5){(c)}
        \put(  37, 60){${\cal T}$}
        \put( 100,  5){(d)}
        \put( 100, 60){${\cal T}(M)$}
    \end{picture}
    \end{center}
    \vspace{-0.25in}
    \caption{The micro-tile $M$ in (a) populates the domain $D$, in (b),
	     $(2 \times 2 \times 2)$ times. $D$ is the domain of
	     trivariate $\cal T$ shown in (c).  Also presented in (c)
	     are the $(2 \times 2 \times 2)$ composed tiles ${\cal
	     T}(M)$.  Finally, (d) shows only the composed tiles ${\cal
	     T}(M)$}
\label{fig-ms-basic}
\end{figure}
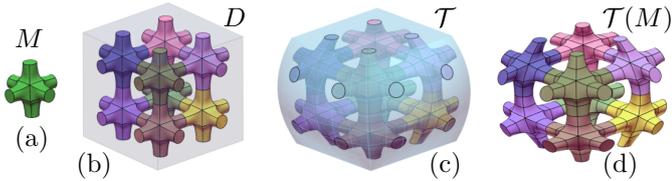

Recently, in~\cite{Elber2016,Massarwi18}, methods and algorithms for
the precise construction of micro-structures using functional
composition~\cite{DeRose1993, Elber1992} were proposed. In that
approach, the design of the macro-shape and the micro-structures of a
porous geometry are decoupled.  A parametric form of a (typically
periodic) micro-tile $M$ is specified as some combination of curves,
surfaces, and/or trivariates while the macro-shape ${\cal T}$ is also
specified as a parametric trivariate function ${\cal T}: D \in
\Reals^3 \rightarrow \Reals^3$.  See Figure~\ref{fig-ms-basic}. $D$,
the domain of ${\cal T}$, is populated with tiles $M_i$, only to
compute the final result as the function composition ${\cal
T}(M_i),~\forall i$.

The approach of~\cite{Elber2016,Massarwi18} has a very simple set of
inputs, namely:
\begin{itemize}
\item A micro-tile $M$, as some combination of parametric
      curves $C(t) = (c_x(t), c_y(t), c_z(t))$, surfaces $S(u, v) =
      (s_x(u, v), s_y(u, v), s_z(u, v))$ and trivariates $T(u, v, w) =
      (t_x(u, v, w), t_y(u, v, w), t_z(u, v, w))$.  Without loss of
      generality, we assume that $M$ is confined to a designated volume,
      specifically a unit cube.  $M$ is typically periodic in the
      sense that the $d_{min}$ faces are $C^0$-continuous with respect
      to $d_{max}$, $d = x, y, z$, and may even be
      $C^k$-continuous, $k > 0$.
\item A trivariate parametric deformation macro-function
      ${\cal T}(x, y, z): D \in \Reals^3 \rightarrow \Reals^3$.
\item $(n_x, n_y, n_z)$: the dimensions of enumerations in ${\cal T}$, 
      in $(x, y, z)$ of the micro-tile $M$.
\end{itemize}
Algorithm~\ref{alg-micro-struct-basic} summarizes this entire process.

\begin{algorithm} 
	\textbf{Input}:\newline
	$M$: a micro-tile consisting of curves and/or surfaces and/or
	     trivariates;\newline
	${\cal T}$: a trivariate parametric deformation macro-function
	     ${\cal T}(x, y, z): D \in \Reals^3 \rightarrow \Reals^3${};
	     \newline
	$(n_x, n_y, n_z)$: the dimensions of the grid enumeration of
        tiles $M$ in the domain of ${\cal T}$;\newline

	\textbf{Output}:\newline
	$\microstruct$: A micro-structure of $(n_x, n_y, n_z)$ tiles in
	a 3D grid deformed to following the macro-shape of ${\cal T}$, via
	functional compositions, as ${\cal T}(M)$;\newline
	
	\textbf{Algorithm}:
	\begin{algorithmic}[1]
	 	\STATE $\microstruct$ := $\emptyset$;
		\FOR {$k = 1, n_z$}
		  \FOR {$j = 1, n_y$}
		    \FOR {$i = 1, n_x$}
		      \STATE $M_{ijk}$ := $M$ positioned at $(i, j, k)$, in $D$;
		      \STATE ${\cal M}_{ijk}$ := ${\cal T}(M_{ijk})$; // via func. composition
		      \STATE $\microstruct$ := $\microstruct \cup \left\{ {\cal M}_{ijk} \right\}$;
		    \ENDFOR
		  \ENDFOR
		\ENDFOR
		\RETURN $\microstruct$;
	\end{algorithmic}
	\caption{\bf Micro-structures synthesis using functional composition}
	\label{alg-micro-struct-basic}
\end{algorithm}

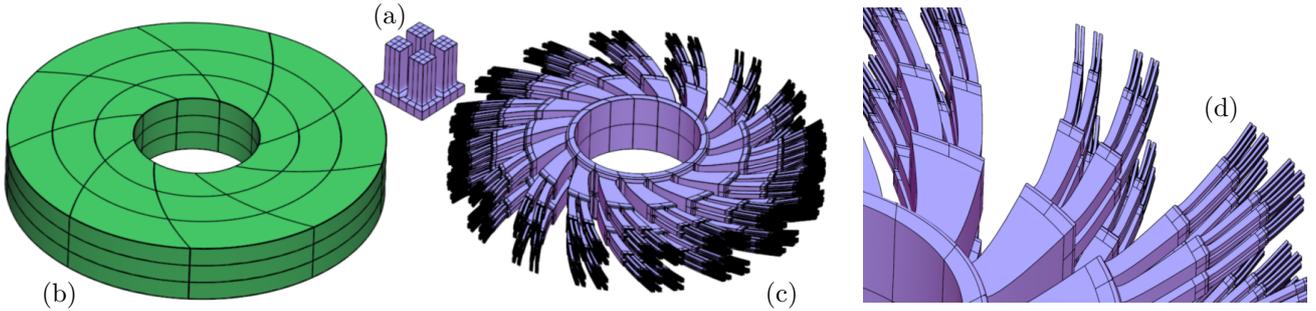
\begin{figure*}
    \begin{center}
        \begin{tabular}{cc}
	    \mbox{\hspace{-0.15in}}
            \epsfig{file=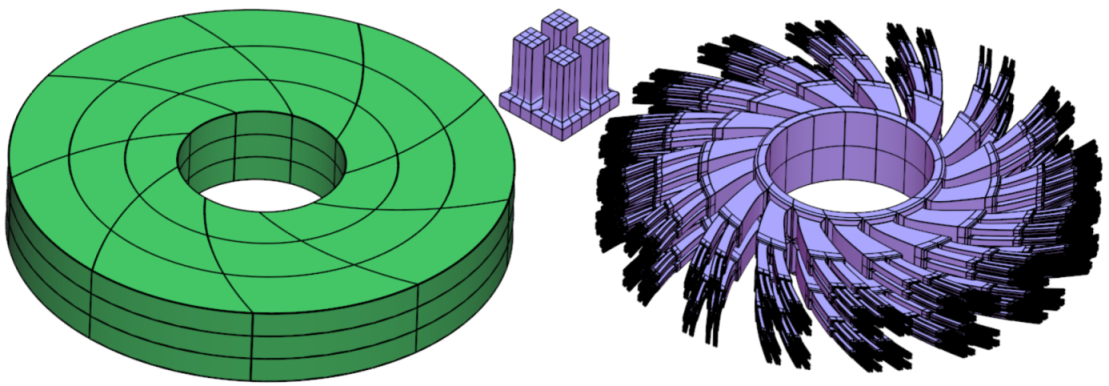,width=4.3in} &
            \epsfig{file=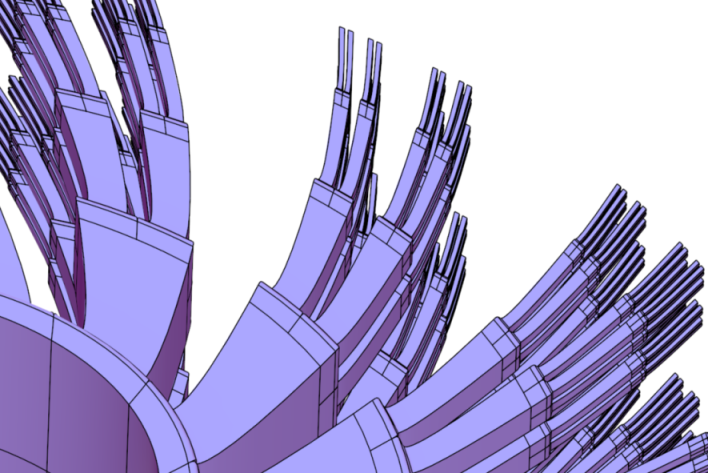,width=2.3in}
        \end{tabular} \\
        \begin{picture}(0,0)
            \put(-109, 110){(a)}
            \put(-233,   5){(b)}
            \put(  38,   5){(c)}
            \put( 202,  75){(d)}
        \end{picture}
    \end{center}
    \vspace{-0.25in}
    \caption{A heat sink design using a hierarchical micro-tile (a)
	    and a macro-shaped twisted ring (b), yielding the heat sink
	    in (c), with large total surface area. (d) shows a zoomed-in
            view on the top right portion of (c).
	    Following the introduction of hierarchical micro-structures
	    in ~\cite{Massarwi18}.}
\label{fig-heat-sink}
\end{figure*}

The result is a precise layout of individual tiles that (continuously)
follows the macro shape.  However, in $D$, the parametric domain of
${\cal T}$, all the placed tiles ${\cal M}_{ijk}$ are of the same
geometry (up to rigid motion).  Clearly, inasmuch as ${\cal T}$ is rarely
isometric, ${\cal T}(M_{ijk})$ can be arbitrarily deformed, which is
where this effort begins.

One application that demonstrates these composition based parametric
micro-structures is the design of a heat sink~\cite{Bornoff2015}.
Following the goal of using maximimal surface area to optimize the 
dissipation of heat while minimizing the volume of material,
the synthesis of the geometry can be achieved with relative ease
through exploiting hierarchical parametric micro-structures.  See
Figure~\ref{fig-heat-sink}.

Striving to control the shapes of the tiles in Euclidean space serves
as a first motivation for this work.  Some general design
specifications can be imposed on the object of design in the form
of strength, weight, heat or electrical conductivity, etc.  One can
then consider the geometry and the topology of individual tiles,
as well as some interior (materials) properties, as degrees of freedom
in some analysis or optimization process, a typical part of a design cycle.

The rest of this work is organized as follows.  In
Section~\ref{sec-prev-work}, previous efforts in synthesizing
micro-structures are surveyed.
Section~\ref{sec-optimal-micro-structs} presents the necessary
building block to enable this proposed design paradigm of
micro-structures. In Section~\ref{sec-examples} some examples
are considered in a variety of engineering applications.  Finally, we
conclude in Section~\ref{sec-conclusion}.


\section{Previous work}
\label{sec-prev-work}

In recent years, the interest in porous geometry and micro-structures,
on one hand, and in heterogeneous materials on the other, has been on
the rise.  One major motivation for 3D representations stems from the
new abilities introduced by additive manufacturing to fabricate such
structures~\cite{Gao2015}.  Porous geometry, micro-structures, and
heterogeneous materials, all enabled by modern additive manufacturing
technologies, are finding applications in a variety of fields from
medicine and biology~\cite{Armillotta2008}, through mechanical and
aero-space engineering to electric engineering~\cite{Rumpf2018} and
material science~\cite{Amadio2015}.

The authors in~\cite{Burczynski2009} perform optimization of
micro-structures towards multi-scale modeling. A set of parameters is
identified at the micro-scale level that govern properties of the model
like the shape, stress, strain, etc.  The finite element method
is used in the analysis step. Evolutionary schemes, which are based
on genetic algorithms, are used to search for optimal parameter values.
Although the \NURB{} representation is used for the shape of the
structures, such an approach lacks precision.

In~\cite{Conde2017} the authors present a framework for modeling
heterogeneous objects using trivariate \Bezier{} patches.  The
\Bezier{} patches have two sets of coordinates. The first set of three
coordinates $(x, y, z)$ prescribe the shape of the object, while the
rest of the coordinates specify the material composition of the
object. While this allows for construction of a wide range of
structures with functionally graded materials (FGM), it only admits a
single level of details, and is limited only to \Bezier{} trivariates.
Modeling of micro-structures, for instance, is beyond the scope
of this work.

The authors in~\cite{Schroeder2005} adopt principles from stochastic
geometry~\cite{Kendall1998} for designing porous artifacts. While such
an approach is suitable for designing random micro-structures, for
example toward bone tissues, it is less favorable for applications
requiring a high level of precision like, for instance, a wing of an
aircraft. Also, the approach does not address the question of ensured
connectivity of the porous structures.  The authors in~\cite{Xiao2016}
use Voronoi tessellations to also generate random porous structures of
three kinds: porous geometries with intersecting fractures,
interconnected tubes and fibers. A set of points is randomly sampled
in the space which gives rise to Voronoi tessellations. Offsets are
then computed for the edges of the tessellations to generate the three
kinds of micro-structures. The authors' aim at fluid flow analysis
while the presented approach is limited to constructing geometry which
is piecewise linear or cylindrical.

In~\cite{Pasko2011} the authors use implicit surfaces for modeling
micro-structures. Their approach allows for design of regular as well
as irregular structures that are amenable to geometric operations
such as blending and deformation. Such an approach does not readily
support the creation of FGM objects. Further, this implicit approach
does not guarantee connectivity in the case of irregular (random)
structures.

In~\cite{Armillotta2008} the authors propose a method for creating
scaffolds as support structures for tissue engineering. The
scaffolds, once fabricated out of some biodegradable or bioresorbable
material, are seeded with (biological) cells and provide support and
shape to tissue during its growth. In this approach, the scaffolds are
modeled as porous micro-structures using a polygonal representation. The
method does not generalize to freeform spline geometry so FGM objects
are not supported.

In~\cite{Wang2005} the authors propose a method for designing
mesoscopic structures using trusses.  The output of the system is in
the form of triangles in STL format. Since this method is tailored to
trusses as the basic building blocks of the structure, the
scope of application is limited. The authors mention filling the volume of
trivariates with tiles, though no details are provided.  The authors
in~\cite{Rosen2008} also propose a framework for designing
additive manufactured mesostructures. Their method is based upon
the process-structure-property-behavior model. The basic building
element is an octet truss which is represented
parametrically. Extensions for support of other types of elements are
not addressed.

In~\cite{Medeiros2015} the authors propose a method to design
cellular structures geared towards additive manufacturing. The
cellular structure is achieved through an adaptive triangulation of
the interior of the solid, with finer tetrahedra along the boundary of
the solid. Their method also supports a dual construction obtained
from the Voronoi diagram of the triangulation. The approach does not
generalize to freeform geometry.

In~\cite{Leblanc2011} the authors propose a modeling primitive based
on a generalized cuboid shape, which is referred to as a block. The
design of a complex object proceeds by laying out blocks, which are
then connected to form the basic shape of the object to be modeled. A
control mesh is then extracted from the faces of the blocks, which
allows parameterization of the surface. While this provides for a
simple and elegant modeling approach, it does not allow the two stage
methodology for the design of micro and macro structures.

Some of the above work either ignores or is not capable of
supporting analysis or optimization applications over the
synthesized geometry.  In others, analysis is feasible while it is
unclear how the results of the analysis or optimization stages can
be fed back into the geometry in order to enhance the designed shape to
complete the design and analysis cycle.  In this work, we propose a
new paradigm for the design of precise FGM micro-structures and porous
geometry using functional composition.  By parametrically controlling
the geometry, topology and materials of individual micro-tiles, one is
provided with a tight link between the design and analysis or
optimization stages, and hence establish a design framework of such
structures.  The next section,
Section~\ref{sec-optimal-micro-structs}, we present the main concept
behind the proposed paradigm.

\section{Micro-structures Synthesis as a Design Paradigm}
\label{sec-optimal-micro-structs}

Recalling Algorithm~\ref{alg-micro-struct-basic} and
following~\cite{Elber2016,Massarwi18}, the construction of
micro-structures involves three main steps:
\begin{enumerate}
\item Specifying the macro shape parametrically.
\item Paving a basic micro-tile $M$ in the domain of the
      macro-shape or the deformation map ${\cal T}$ as
      $\left\{M_{ijk}\right\}$.
\item Functionally composing the tiling created in step 1,
      $\left\{M_{ijk}\right\}$, into the deformation map ${\cal T}$,
      to obtain the required micro-structure ${\cal T}(M_{ijk})$.
\end{enumerate}

Assuming the tile $M$ is a (trimmed) trivariate or a set of such
(trimmed) trivariates, and following this pavement process,
isogeometric analysis can be immediately applied to the structure, as
is done in ~\cite{Massarwi18}, for example.  Then, the results of the
analysis and/or optimizations can be fed directly to the next geometry
synthesis cycle in which the parameters of the geometry/material
properties, etc., are modified to follow the provided constraints and
analysis or optimizations.  In order to specify material properties
and other fields, the control points for the trivariates are extended
to $\Reals^k$, $k \ge 3$.  The first three dimensions are used for the
geometry, whereas higher dimensions are optionally used for holding
scalar, vector and tensor properties, such as materials.

\begin{figure}
    \begin{center}
        \begin{tabular}{c}
	    \mbox{\hspace{-0.15in}}
            \epsfig{file=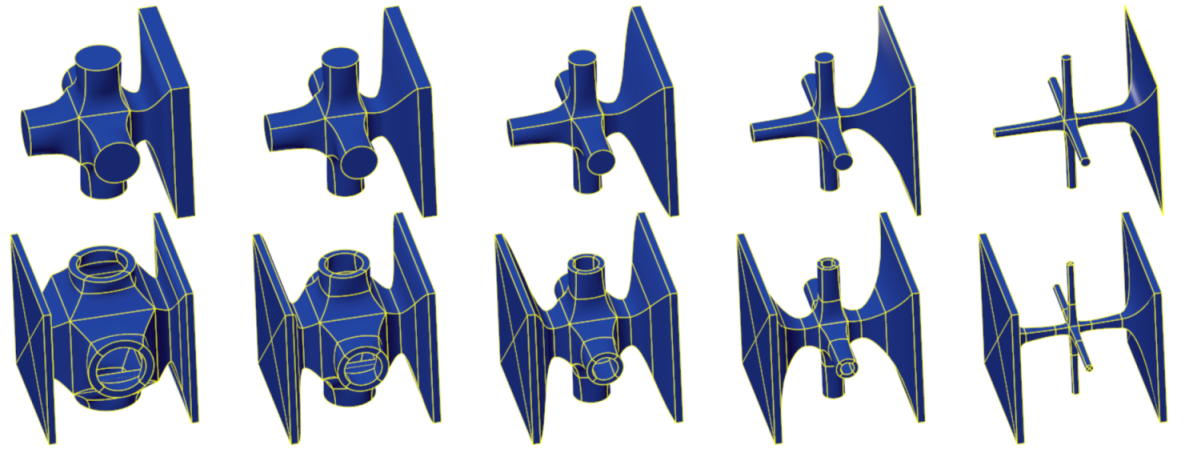,width=3.5in}
        \end{tabular}
    \end{center}
    \vspace{-0.25in}
    \caption{Parametrically varying wall thicknesses/tubes diameters in a
	     solid tile $M$ with a single boundary wall on its right side
	     (top).
	     Parametrically varying diameters of hollowed tubes in a tile
  	     $M$ with a pair of fixed thickness boundary walls (bottom).}
\label{fig-parametric-tile-geom}
\end{figure}

Let ${\cal P} = \left\{p_1, p_2,...,p_n\right\}$ be a set of $n$
parameters for a family of geometric shapes, topologies and FGM
distributions, etc., of a tile $M$.  Typically, these parameters will
be constrained so critical properties such as the sign of the
Jacobian, will not be affected.  Figure~\ref{fig-parametric-tile-geom}
shows different tiles with parametric control over the wall
thicknesses of both solid and hollow $M$, and illustrates parametric
control over the diameters of the tubes in a tile.

Algorithm~\ref{alg-micro-struct-basic} can now be modified to
accommodate {\em black box} optimizations or analysis tools to
manipulate the micro-structure.  See
Algorithm~\ref{alg-micro-struct-optimal}.

\begin{algorithm} 
	\textbf{Input}:\newline
	$M({\cal P})$: a {\em parametric} micro-tile consisting of curves
	    and/or surfaces and/or trivariates;\newline
	${\cal T}$: a trivariate parametric deformation macro-function
	     ${\cal T}(x, y, z): D \in \Reals^3 \rightarrow \Reals^3$;
	     \newline
	$(n_x, n_y, n_z)$: the initial dimensions of the grid
	     enumerating tiles $M$ in the domain of ${\cal T}$;
	     \newline
	${\cal O}$: an optimizer - a {\em black box} that optimizes the
	     synthesized micro-structure;\newline

	\textbf{Output}:\newline
	$\microstruct$: A micro-structure of $(n_x, n_y, n_z)$ tiles
	deformed to following the macro-shape of ${\cal T}$ via
	functional composition ${\cal T}(M({\cal P}))$ as influenced by 
	optimizer ${\cal O}$;\newline
	
	\textbf{Algorithm}:
	\begin{algorithmic}[1]
	      \STATE ${\cal P}$ := init parameters for $\microstruct$;
	      \WHILE {Not Optimized}
	 	\STATE $\microstruct$ := $\emptyset$;
		\FOR {$i = 1, n_x$}
		  \FOR {$j = 1, n_y$}
		    \FOR {$k = 1, n_z$}
		      \STATE ${\cal P}_{ijk}$ := geometry/topology/material parameters of tile $ijk$, as prescribed by ${\cal P}$;
		      \STATE $M_{ijk}$ := $M({\cal P}_{ijk})$ at position $(i, j, k)$, in $D$;
		      \STATE ${\cal M}_{ijk}$ := ${\cal T}(M_{ijk})$; // via func. composition
		      \STATE ${\microstruct}$ := ${\microstruct} \cup \left\{ {\cal M}_{ijk} \right\}$;
		    \ENDFOR
		  \ENDFOR
		\ENDFOR
		\STATE ${\cal P}$ := solve ${\cal O}({\microstruct})$;
	     \ENDWHILE
	     \RETURN ${\microstruct}$;
	\end{algorithmic}
	\caption{\bf Micro-structures synthesis using functional composition and optimization}
	\label{alg-micro-struct-optimal}
\end{algorithm}

In Algorithm~\ref{alg-micro-struct-optimal}, the geometry is
synthesized using a multi-parameters family of micro-tiles.  The
optimizer ${\cal O}$ sets the parameters (i.e. wall thickness or pipe
diameter) at iteration $i$, to affect only the synthesized geometry at
iteration $i+1$.  While Algorithm~\ref{alg-micro-struct-optimal}
hints at parameters that control the geometry, the parameters
can also control material properties like heat conductivity.

For the first pass of the optimization,
Algorithm~\ref{alg-micro-struct-optimal} needs to synthesize an
initial micro-structure so the optimizer ${\cal O}$ has something on
which to operate.  This initial micro-structure is application
dependent and in Section~\ref{sec-examples} we present several such
applications.

The presented scheme provides for significant freedom to modify
individual micro-tiles while preserving the continuity of the the
entire arrangement.  If a tile is modified in the domain of the
deformation macro-function, and its adjacent micro-tiles are updated
accordingly to preserve the continuity, the outcome will preserve
continuity as well (assuming the deformation macro-function is
sufficiently continuous).  Further, the continuity of the geometry
must be preserved as well as the continuity of properties such as
material.  Finally, the modified topology should not affect the
continuity.  For example, it is infeasible to modify a branching tile
from 2 branches to 4 branches without also modifying adjacent tiles to
retain continuity properties.

\section{Applications and Examples}
\label{sec-examples}

In this section we present four different design optimization
application of micro-structures.  In
Section~\ref{subsec-examples-heat-exchange} micro-structures are
optimized toward heat exchangers.
Section~\ref{subsec-examples-rocket-fuel} considers the question of
solid heterogeneous rocket fuel design.  In
Section~\ref{subsec-examples=wing-design} a wing design based on
micro-structures is considered, and finally, in
Section~\ref{subsec-examples-heating-extruders}, local thermal
control, is optimized in plastic extruders.

\subsection{Heat Exchanger}
\label{subsec-examples-heat-exchange}
Although often not visible from the exterior of a mechanical system
and not the focus of widespread attention, heat exchangers are used
universally for heating and cooling functions. Common examples of heat
exchangers include radiators, oil coolers, refrigeration/air
conditioners/heat pumps, and etc. They occur in a variety of types
whose forms are typically driven by applications and traditional
manufacturing techniques. 
 
 Typically, heat exchangers are selected
from a catalog of rectilinear or cylindrical prismatic shapes, thus
one is forced to design around the existing standard pre-manufactured
shapes. This can lead to awkward, inefficient applications because the
design must be adapted to conform to and accommodate the catalog
available heat exchanger shapes. The designer is required to adjust
the flow so that the conditions at the inlet of the heat exchanger can
result in good heat transfer. Additive manufacturing opens the
opportunity to design custom heat exchangers specific to the cavity or
envelope that naturally occurs in an emerging design. 

\begin{figure}
    \begin{center}
        \epsfig{file=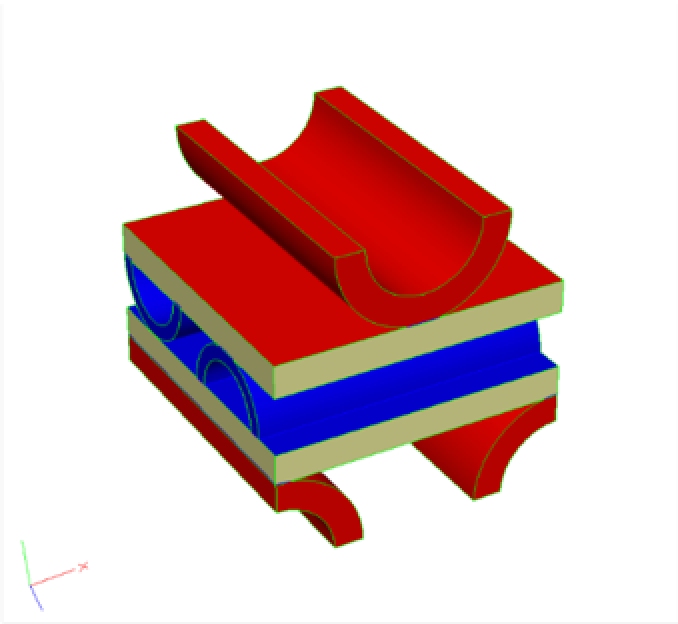,width=1.70in}
         \epsfig{file=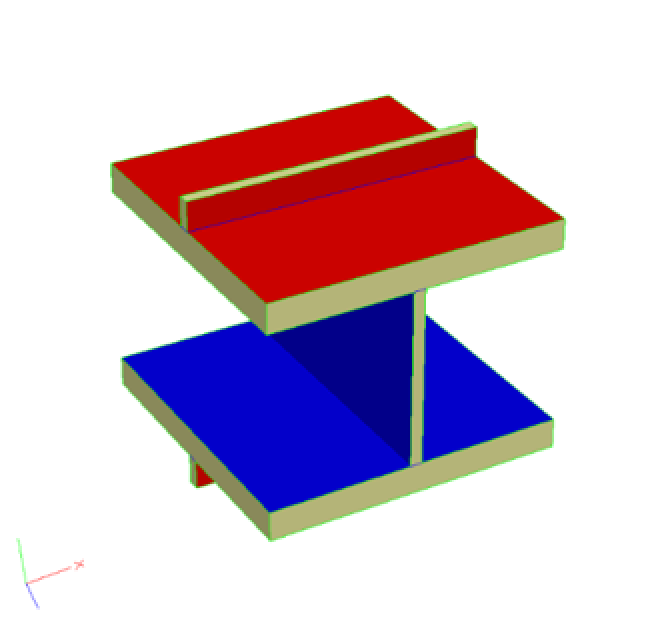,width=1.70in}
    \end{center}
    \vspace{-0.25in}
    \caption{This contrasts a traditionally used element on the left with a simpler tile element on the right that can be easily generated via additive manufacturing. Note that the red and blue colored surfaces represent those whetted to the individual fluid channels, where the tan color refers to metal.}
\label{fig-HX-tiles}
\end{figure}

The simplest, inexpensive and most common heat exchangers are
manufactured by taking a sandwich of corrugated metal, a metal sheet
and then corrugated metal (at 90 degrees to the first corrugated
layer), another sheet and then repeating. This suite of plies is then
welded together and {\it headers} added to create a closed loop for
one of the fluids. If one were to generate a single tile of this
device it would look like the left-hand image seen in
Figure~\ref{fig-HX-tiles}. For the rest of the discussion we will be
using a simpler repeat pattern that is easily constructed by additive
manufacturing and can be seen on the right side of
Figure~\ref{fig-HX-tiles}.

\begin{figure}
    \begin{center}
        \epsfig{file=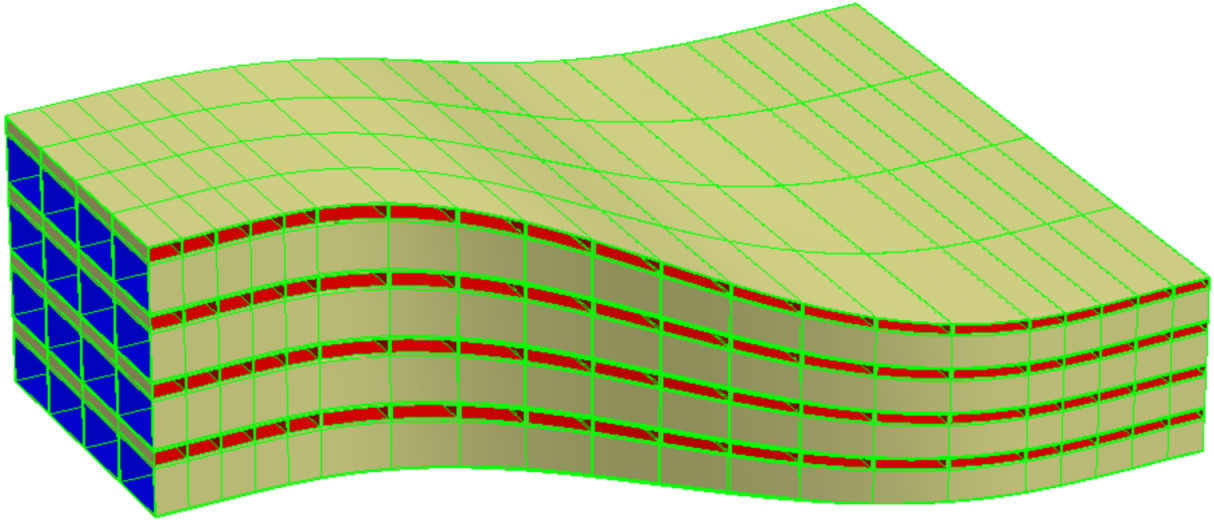,width=1.70in}
        \epsfig{file=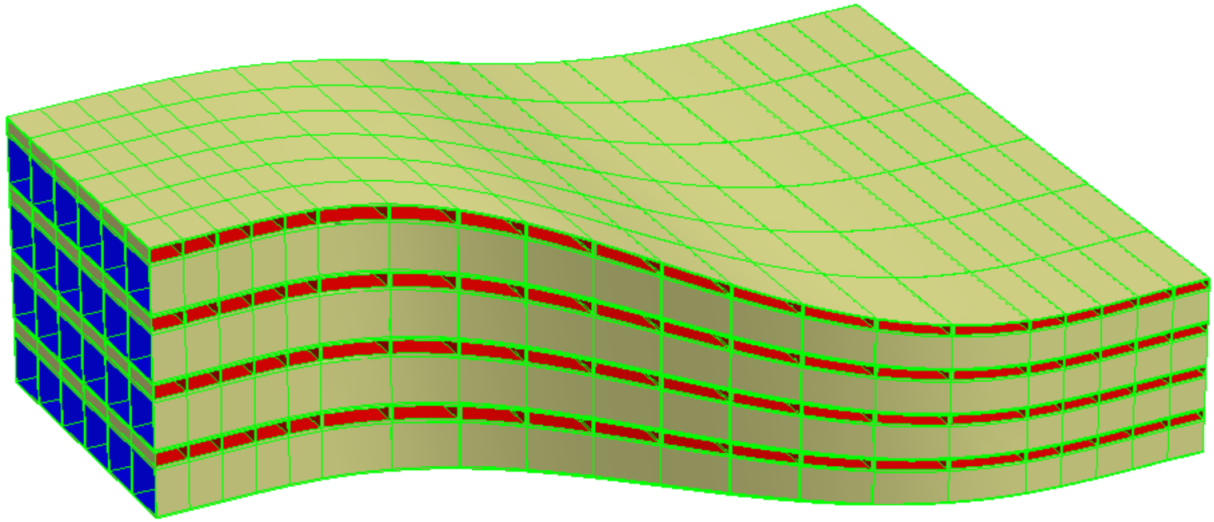,width=1.70in}
        \epsfig{file=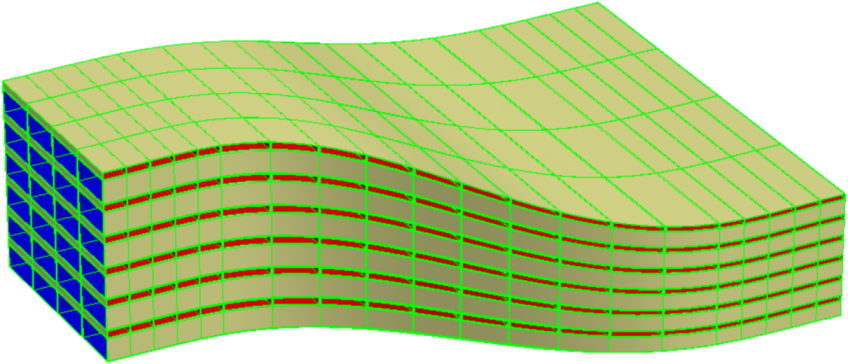,width=1.70in}
        \epsfig{file=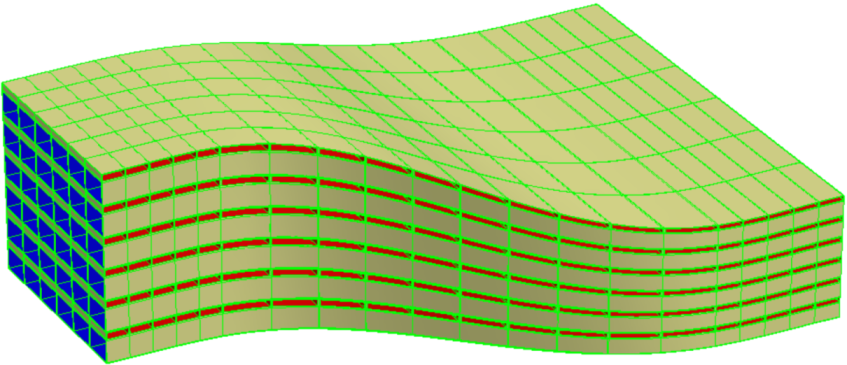,width=1.70in}
    \end{center}
    \vspace{-0.25in}
    \caption{This sequence of images illustrates the flexibility afforded by a parametric heat exchanger design. The entire available cavity or duct can be used for the device and the number of passages of either fluid can be adjusted (as well as the spacing). The upper left image is the baseline where the upper right has increased the number of {\it blue} passages. The lower left image show the same duct with a larger number of {\it red} passages. The lower right image is simply more passages for both fluids.}
\label{fig-HX-ducts}
\end{figure}

Figure~\ref{fig-HX-ducts} depicts the composition of the tile seen in
the right side of Figure~\ref{fig-HX-tiles} into a duct with changing
cross-sections. In this case the spacings in the duct are changing,
which is driven by the the knot sequence of the geometry that defines
the duct (${\cal T}_{orig}$).

An effective heat exchanger maximizes the heat transfer over the
entire device. This is not an isolated geometric problem but is a
complex heat transfer problem relating the geometry, metal properties,
thermal gradients and flow characteristics of the device. A low
fidelity, lumped-parameter heat exchanger model has been constructed
for this problem that can output trivariate metal thicknesses,
hot-to-cold area ratios, and relative tile sizes throughout ${\cal
T}_{orig}$. The model takes into consideration the physical laws in
each hot and cold channel (including temperature and pressure of the
appropriate fluid), deals with mass and energy conservation while
using a convective heat transfer model. This is accomplished before
the geometry of the heat exchanger is composed.

The first step in building the device is to reconstruct the duct so
that its knot sequence reflects the relative spacings of the tiles
($\microstruct$) in physical space (${\cal T}_{orig} \Rightarrow {\cal
T}$). This is done so that each tile is generated with the appropriate
size. Algorithm~\ref{alg-HX} is then followed, where it should be
noted that each tile is individually constructed during the
composition. The tile's parameters are sampled from the scalar
trivariate fields, so that after the construction is complete the
geometry matches up at the individual tile interfaces. This allows for
the building of a coherent structure (by sewing) that can be
represented as a geometric {\it solid}.

\begin{algorithm}
	\textbf{Input}:\newline
	$M({\cal P})$: a {\em parametric} micro-tile consisting of the canonical {\it solid} shape;\newline
	${\cal T}$: a trivariate parametric deformation macro-function; \newline
	${\cal H}$: a trivariate scalar of hot-to-cold area ratios; \newline
	${\cal K}$: a trivariate scalar of metal thicknesses; \newline
	$(n_x, n_y, n_z)$: the initial dimensions of the grid
	     enumerating tiles $M$ in the domain of ${\cal T}$;
	     \newline

	\textbf{Output}:\newline
	$\microstruct$: A micro-structure of $(n_x, n_y, n_z)$ tiles
	deformed to following the macro-shape of ${\cal T}$ via
	functional composition ${\cal T}(M({\cal P}))$ as influenced by 
	the sizing from the lumped-parameter heat exchanger model;\newline
	
	\textbf{Algorithm}:
	\begin{algorithmic}[1]
	 	\STATE $\microstruct$ := $\emptyset$;
		\FOR {$i = 1, n_x$}
		  \FOR {$j = 1, n_y$}
		    \FOR {$k = 1, n_z$}
		      \STATE ${\cal H}_{ijkm}$ := hot-to-cold area ratio at the 8 corners of the tile ($m = 1, 8$);
		      \STATE ${\cal K}_{ijkm}$ := metal thicknesses at the 8 corners of the tile ($m = 1, 8$);
		      \STATE ${\cal K}^{-1}_{ijkm}$ := ${\cal T}^{-1}({\cal K}_{ijkm})$, the deformed metal thicknesses so that after composition the correct physical thicknesses are realized;
		      \STATE ${\cal P}_{ijk}$ := geometry/topology parameters of tile $ijk$, which is a function of ${\cal H}_{ijkm}$ and
		      ${\cal K}^{-1}_{ijkm}$;
		      \STATE $M_{ijk}$ := $M({\cal P}_{ijk})$ at position $(i, j, k)$;
		      \STATE ${\cal M}_{ijk}$ := ${\cal T}(M_{ijk})$; // via func. composition
		      \STATE ${\microstruct}$ := ${\microstruct} \cup \left\{ {\cal M}_{ijk} \right\}$;
		    \ENDFOR
		  \ENDFOR
		\ENDFOR
	     \RETURN Sew(${\microstruct}$);
	\end{algorithmic}
	\caption{\bf Micro-structures synthesis for the heat exchanger}
	\label{alg-HX}
\end{algorithm}

\begin{figure*}
    \begin{center}
            \epsfig{file=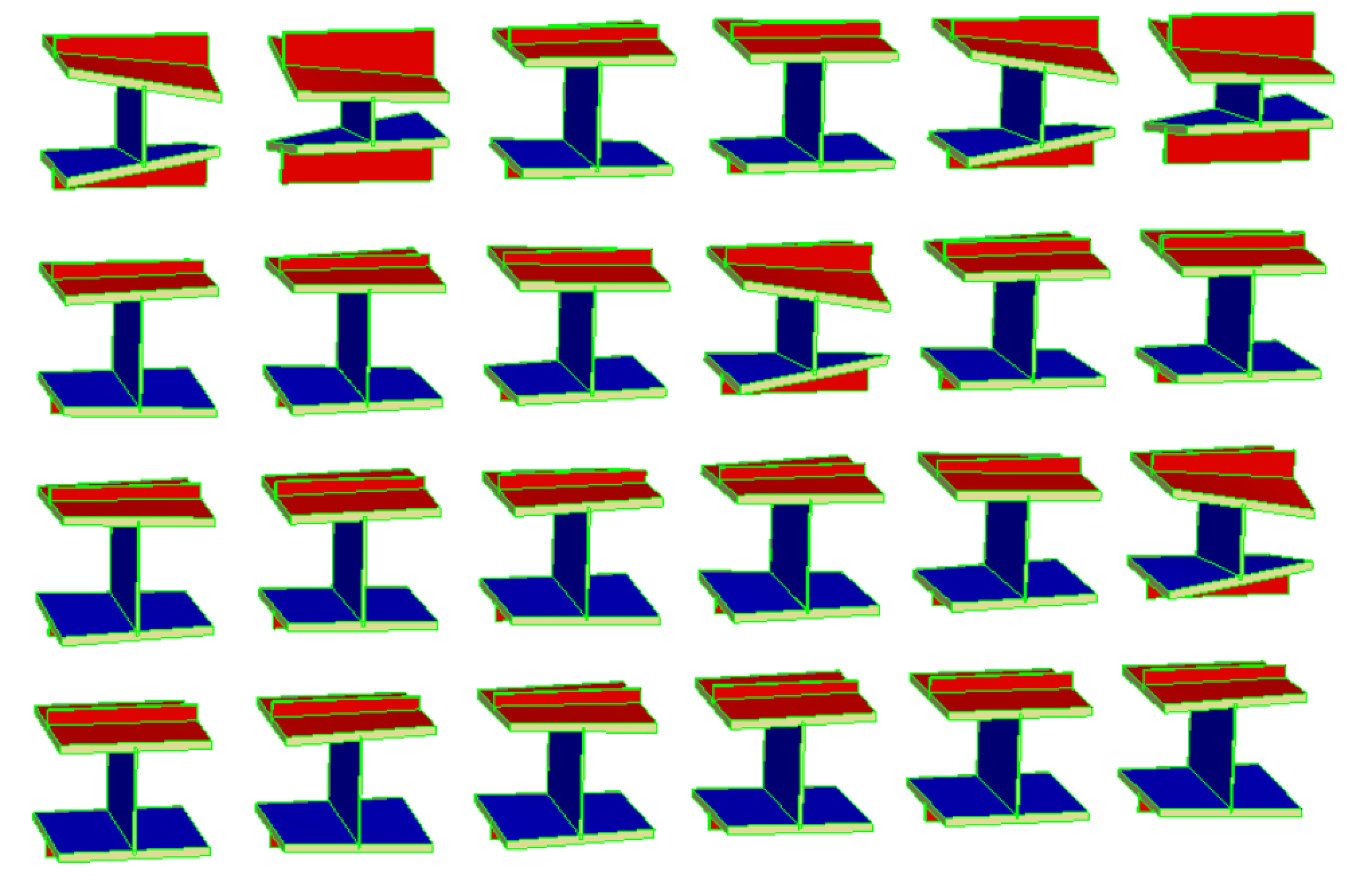,width=2.8in}
            \epsfig{file=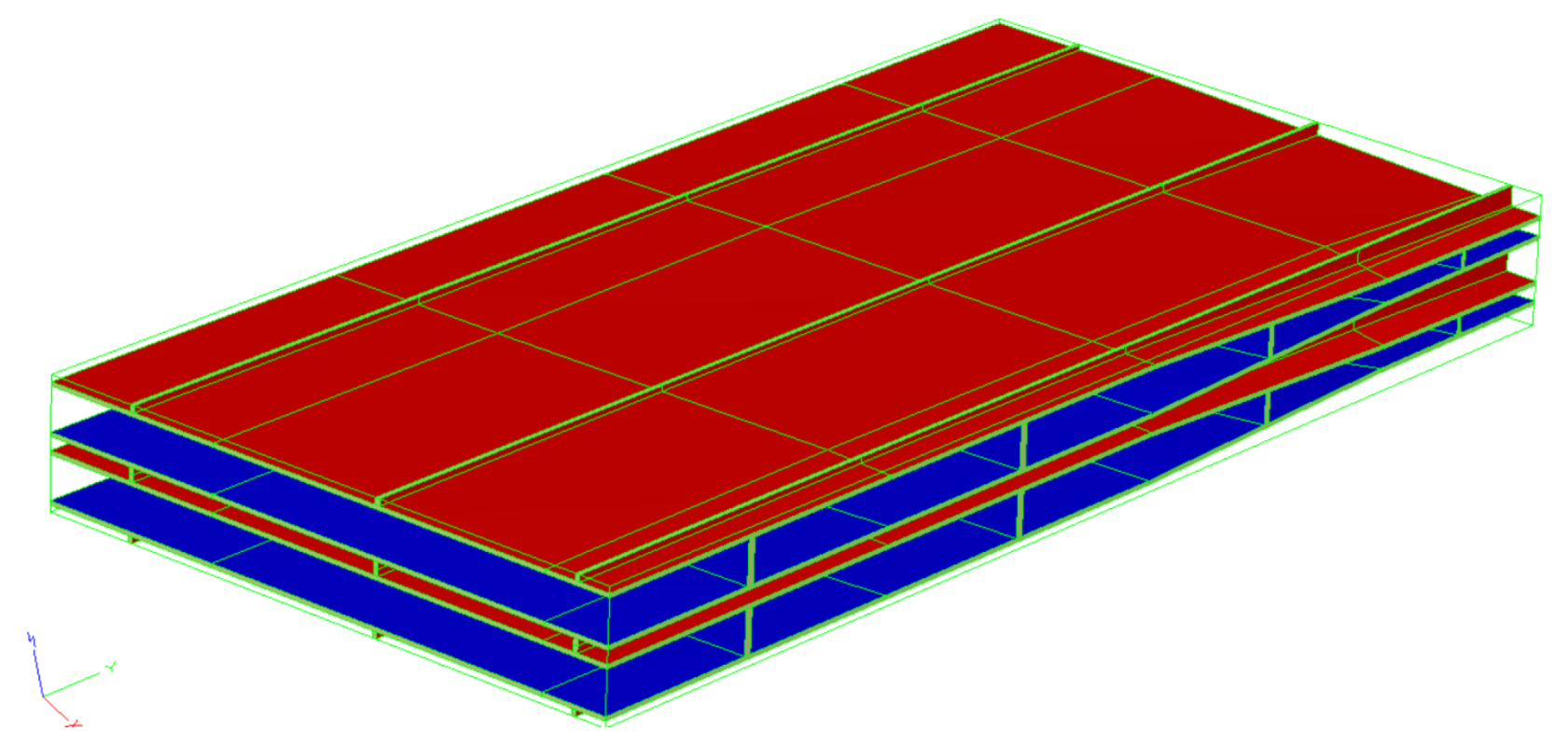,width=3.8in}
    \end{center}
    \vspace{-0.25in}
    \caption{The left side depicts the individual parametric tiles in their pre-deformed unit representation. The right side image shows the composed resulting heat exchanger in physical space.}
\label{fig-HX-compose}
\end{figure*}

Figure~\ref{fig-HX-compose} shows the results of Algorithm~\ref{alg-HX} on a simple rectilinear outer shape with a small number of tiles. The left side of the figure show all of the individual tiles used in the composition (${\cal M}_{ijk}$), where the right-hand side of Figure~\ref{fig-HX-compose} shows the composed result driven by the spacings and thicknesses output from the low fidelity, lumped-parameter heat exchanger model.

\subsection{Design of a Solid Fuel Rocket Grain}
\label{subsec-examples-rocket-fuel}

This application concerns the use of accelerants and retardants mixed
with the propellant in a solid rocket engine. Mixing the fuel is
considered in order to control and plan the burn rate and thereby the
expected thrust. The problem given is to match a prescribed thrust
profile and the geometry of the solid fuel rocket casing, which
defines the end state of the burn.

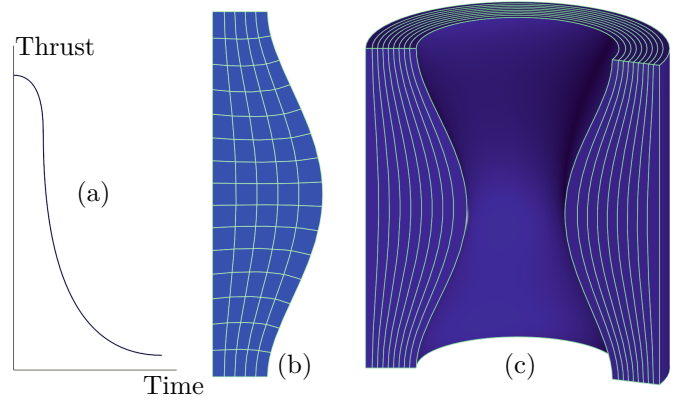
\begin{figure}
    \begin{center}
    \begin{tabular}{c}
	\mbox{\hspace{-0.15in}}
        \epsfig{file=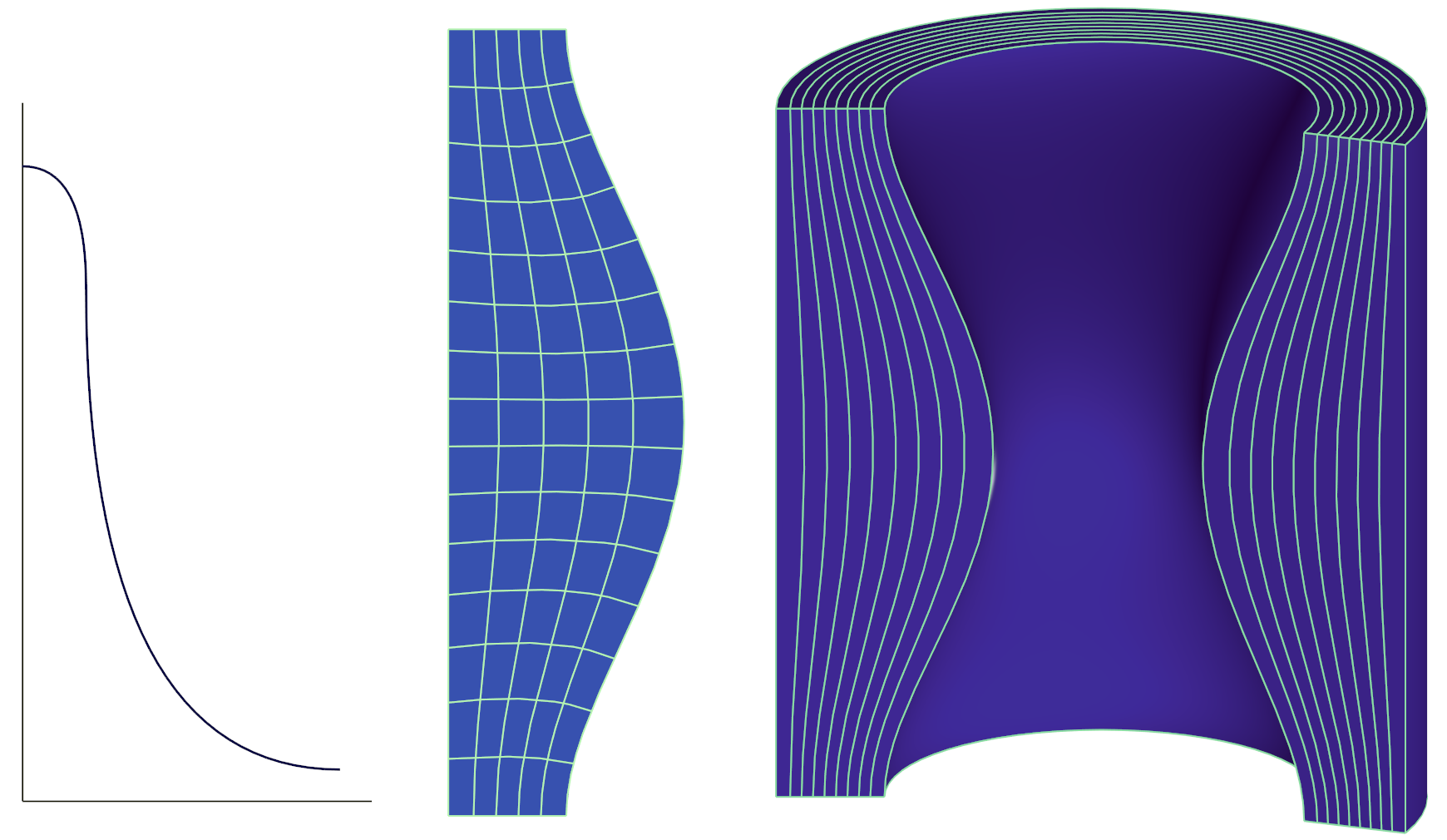,width=3.5in}
    \end{tabular}
    \begin{picture}(0,0)
        \put(-100, 75){(a)}
        \put( -25, 10){(b)}
        \put(  60, 10){(c)}
        \put(-75,   2){Time}
        \put(-123, 130){Thrust}
    \end{picture}
    \end{center}
    \vspace{-0.25in}
    \caption{The thrust profile (a) is provided along with the rocket fuel
	     cross section (b), and an orthogonal (conformal)
	     parametrization.  Rotating (b) defines a
	     volume-of-revolution, shown in (c), that specifies the
	     solid fuel grain.  Also shown in (c) is the division of
	     the volume-of-revolution into layers.}
\label{fig-rocket-fuel-basic}
\end{figure}

For simplicity and to begin with, we assume that the micro-structure
elements are solid and that the geometry of the solid fuel grain is a
volume-of-revolution constructed by rotating a 2D planar section about
the rocket's longitudinal axis, not an atypical setting. See
Figure~\ref{fig-rocket-fuel-basic}.  In other words, each
micro-structure element is a full, possibly heterogeneous, logical cuboid.

The micro-structures-based construction scheme of the heterogeneous solid
fuel grain follows Algorithm~\ref{alg-rocket-fuel-optimal}.

\begin{algorithm} 
	\textbf{Input}:\newline
	${\cal S}(u, w)$: a 2D section of
		rocket fuel geometry, $w$ being a (orthogonal)
		parametrization of the fuel, inside out;\newline
	$T(t),~t \in [0,1]$: a thrust profile function, over time;\newline
	$n$:    micro-structure sampling rate;

	\textbf{Output}:\newline
	$\microstruct$: A micro-structure with heterogeneous composition
		${\cal AR}$ (accelerants/retardants)
		of a 3D volume-of-revolution shaped fuel grain ${\cal V}$, with
	        ${\cal S}$ as its cross section, satisfying
		the thrust profile $T(t)$;
	
	\textbf{Algorithm}:
	\begin{algorithmic}[1]
	    \STATE ${\cal V}(u, v, w)$ := a volume-of-revolution
			trivariate defined by rotating section ${\cal S}(u, w)$;
	    \STATE ${\cal V}_k(u, v, w), ~k = 1, n$ := a partitioning in $w$ of
			${\cal V}(u, v, w)$ into $n$ trivariate layers;
	    \FOR {$k = 1, n$}
		\STATE $TotalLayerThrust$ := 0;
	        \FOR {$i = 1, n$}
		    \FOR {$j = 1, n$}
		        \STATE ${\cal M}_{ijk}$ := Tile $ij$ in layer ${\cal V}_k$;
			\STATE $A_{ijk}$ := $TileFrontArea({\cal M}_{ijk})$; // in $uv$
			\STATE $d_{ijk}$ := $TileDepth({\cal M}_{ijk})$; // in $w$;
			\STATE $Thrust_{ijk}$ := $A_{ijk} d_{ijk}$;
			\STATE $TotalLayerThrust$ += $Thrust_{ijk}$;
		    \ENDFOR
		\ENDFOR

		\STATE $d_k$ := average depth of layer ${\cal V}_k(u, v, w)$;
		\STATE $ThrustRatio$ := $T(k/n) / TotalLayerThrust$;
	        \FOR {$i = 1, n$}
		    \FOR {$j = 1, n$}
			\STATE ${\cal AR}_{ijk}$ := $ThrustRatio * (d_{ijk} / d_k)$;
			\STATE $\microstruct$ := $\microstruct \cup \left\{ {\cal M}_{ijk}({\cal AR}_{ijk}) \right\}$;
		    \ENDFOR
		\ENDFOR
	     \ENDFOR
	\end{algorithmic}
	\caption{\bf Heterogeneous rocket fuel optimization}
	\label{alg-rocket-fuel-optimal}
\end{algorithm}

Algorithm~\ref{alg-rocket-fuel-optimal} builds the
volume-of-revolution ${\cal V}(u, v, w)$ (in line 1) by rotating
${\cal S}(u, v)$ about the rocket's longitudinal axis, only to
partition this trivariate (in line 2) to $n$ layers ${\cal V}_k(u, v,
w)$, by subdividing ${\cal V}$ at $n-1$ different $w$ values.  Note
that each ${\cal V}_k(u, v, w)$ is a trivariate as well - see
Figure~\ref{fig-rocket-fuel-basic}~(c).  Each layer ${\cal V}_k$ has a
varying thickness.  Further, the front surface (burning) area of
${\cal V}_k$ varies as well, as the burning progresses in the layer.
In the limit, however, as $n$ goes up and the thicknesses of the
layers are vanishing, the front surface area can be assumed fixed.  We
plan the burning rate (using accelerants/retardants) so each layer
will burn completely before the next layer starts to burn.  That is,
as the burning front progresses, it simultaneously interpolates the
next layer.  Thus the layers serve literally as synchronization
surfaces for the timing of the burn front progression.  Then, in line
7-11 of Algorithm~\ref{alg-rocket-fuel-optimal}, we compute for each
tile ${\cal M}_{ijk}$ in layer ${\cal V}_k$, its burn front area (in
$uv$) and depth (in $w$) and estimate ${\cal M}_{ijk}$'s total thrust
as its front area times its depth.

\begin{figure*}
    \begin{center}
    \begin{tabular}{c}
	\mbox{\hspace{-0.15in}}
        \epsfig{file=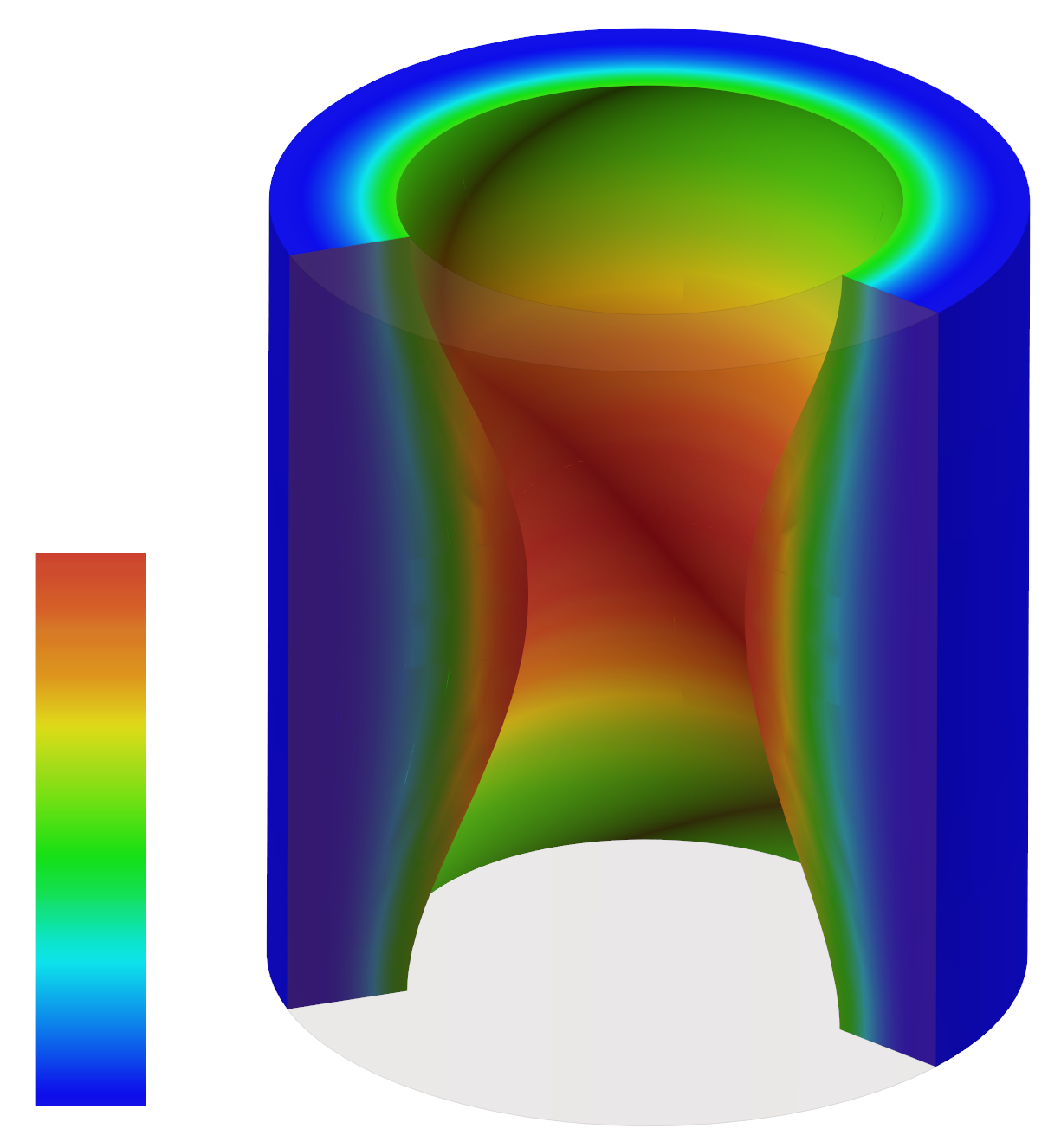,height=1.8in}
        \epsfig{file=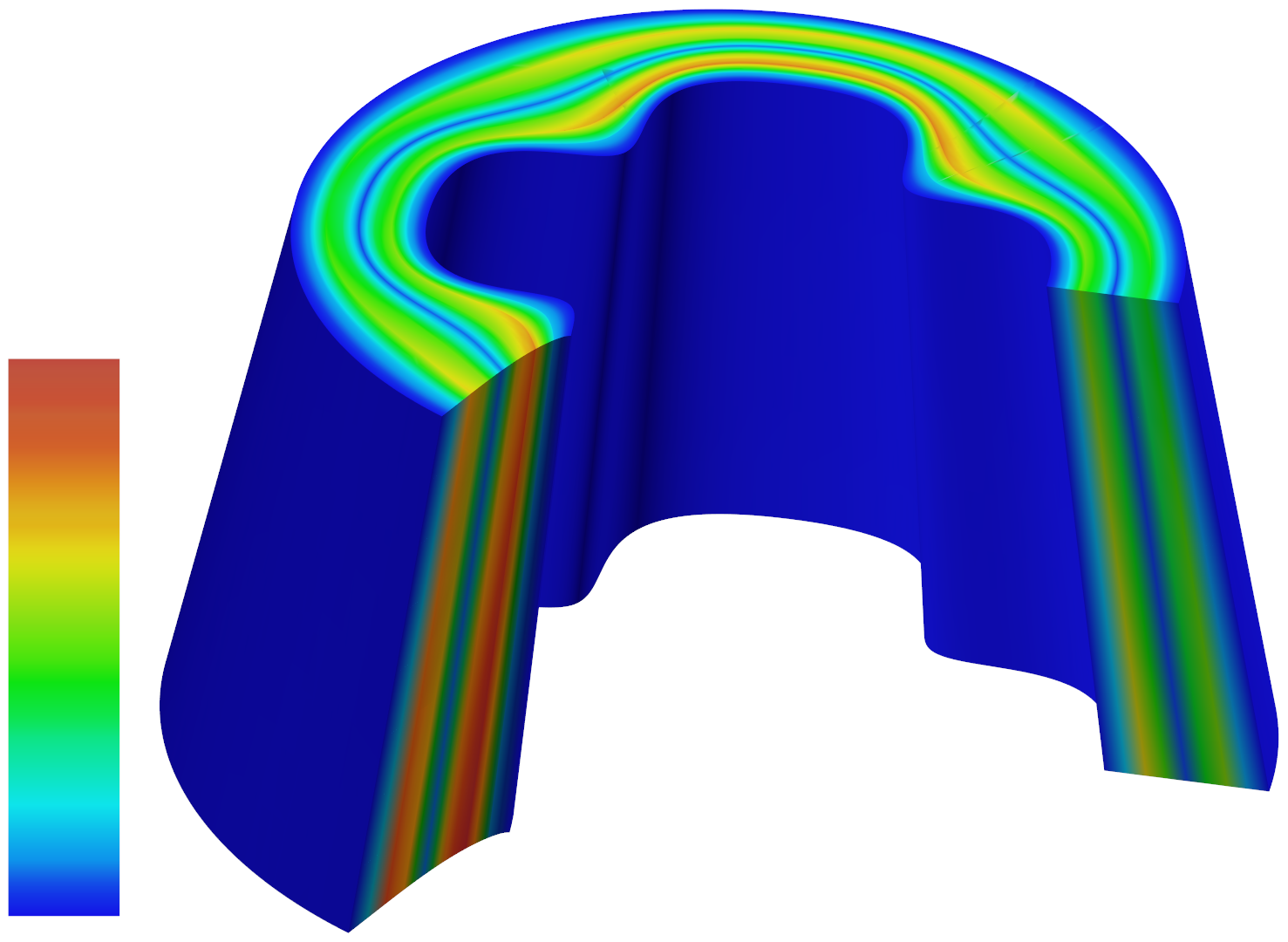,height=1.8in}
        \epsfig{file=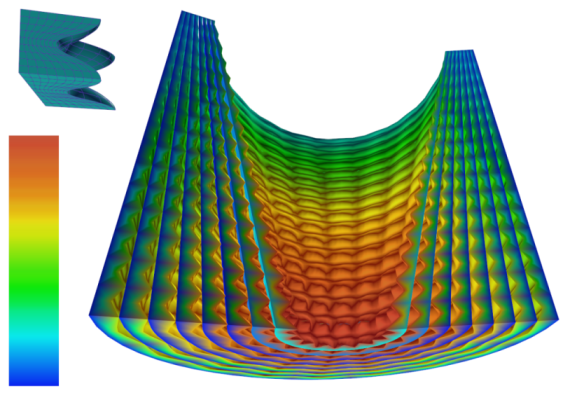,height=1.8in}
    \end{tabular}
    \begin{picture}(0,0)
        \put(-490, 45){(a)}
        \put(-280,-35){(b)}
        \put( -25, 20){(c)}
    \end{picture}
    \end{center}
    \vspace{-0.05in}
    \caption{The thrust profile and geometry from
	     Figure~\protect\ref{fig-rocket-fuel-basic} yields the (sliced)
             result shown in (a), and using
	     Algorithm~\ref{alg-rocket-fuel-optimal}.
	     Red denotes augmentation with accelerants,
	     and blue, the addition of retardants.  Note the high
	     (red) burning rate at the interior middle surface, in an aim to
	     achieve the specified high initial thrust, only to
	     reduce into all blue (retardants) as time advances.
	     In (b), one example of a non volume-of-revolution solid fuel
	     is presented, with a thrust profile having two
	     peaks. Finally, (c) portrays the possibility of using
	     non-solid fuel tile (tile shown on top left in (c)),
	     in an effort to further enhance the burning rate and
	     hence, the thrust. In (c), the thrust is a constant curve (while
	     the bottom is burning faster because it is thicker).}
\label{fig-rocket-fuel-basic-result}
\end{figure*}
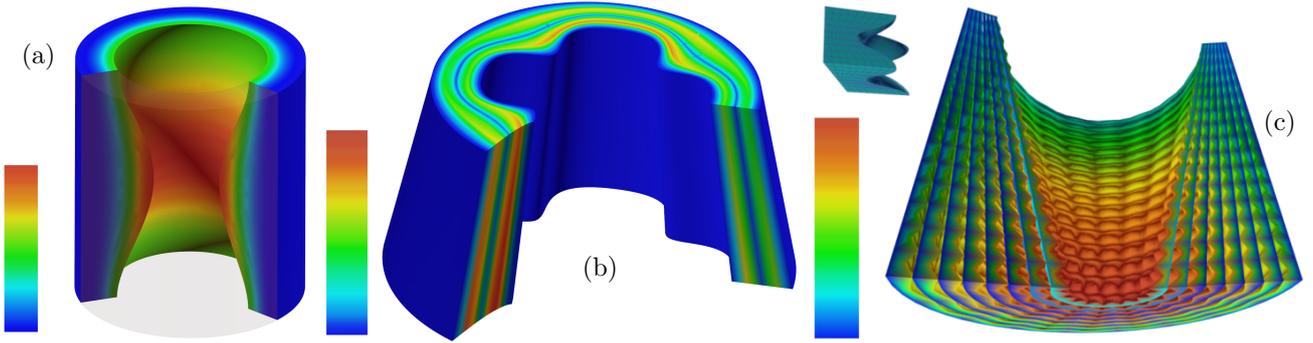

Accumulating the total thrust that a layer produces as
$TotalLayerThrust$, we globally normalize, in line 15, the relative
amount of accelerant $({\cal AR}_{ijk} >1)$ or retardant $({\cal
AR}_{ijk} < 1)$ that is required for this layer (time step).  This
global normalization depends on the ratio between what is the desired
thrust at this time step $T(k/n)$ and the basic thrust produced in
the layer $TotalLayerThrust$.  Then, in line 18, the global
normalization is combined with a local, tile-level, normalization.
For tile ${\cal M}_{ijk}$, the relative depth $d_{ijk}$ with respect to
the layer's average depth sets the local normalization.  The deeper
the tile, the faster it must burn (and hence requires more
accelerant) to ensure that the entire layer burns out simultaneously. 
Then the burn front interpolates the next layer synchronously in time. 

Figure~\ref{fig-rocket-fuel-basic-result}~(a) shows one simple result,
using the thrust profile and geometry from
Figure~\protect\ref{fig-rocket-fuel-basic}.
Figure~\ref{fig-rocket-fuel-basic-result}~(b) shows that designing a fuel grain which is not a
volume-of-revolution solid fuel is feasible as well. Finally,
Figure~\ref{fig-rocket-fuel-basic-result}~(c), shows the potential use
of porous tiles, in an aim to further increase the burning surface
area and hence the thrust.

\subsection{Wing Design}
\label{subsec-examples=wing-design}

Aircraft wings have many functions and their design and packing
reflects this complexity. Wings primarily provide lift (due to their
outer shape), store fuel, contain mechanisms to provide additional
lift during maneuvers such as take-off and landings, hold landing gear
and the mechanisms to both extend and retract, and provide the ability
to steer the craft through the use of flaps. This means that the
internals of wings contain housings for flaps, slats, landing gear,
fuel tank(s), and all of the wiring and pneumatic piping (both primary
and backup) required for the full functioning of the aircraft. The
wing must be able to withstand the forces and stresses encountered
during the mission and therefore the internal structures that support
the wing must be flexible, robust and light in weight.

Traditionally, manufacturing has given us wing internal structures
that are a collection of a small number of {\it spars} (usually 2) and
a handful of {\it ribs} that are usually orthogonal and are all
interconnected.  The {\it spars} and a couple of {\it ribs} can form a
{\it wing box} that may hold fuel. Usually, the {\it ribs} have many
cutouts to lighten the structure and allow for the running of wiring
and piping.  Overall, the skin (at least between the {\it spars} and
the outermost {\it rib}) is part of the structure (that is, the skin
carries load).  At times the expanse of skin (partitioned into {\it
bays}) allows buckling (not good) and therefore may also require {\it
stiffeners} to reinforce the structure in appropriate regions.
%

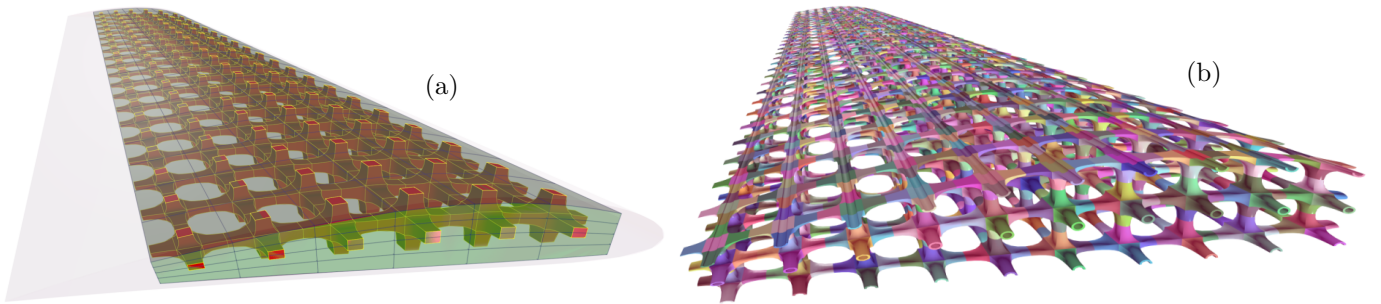
\begin{figure*}
    \begin{center}
    \begin{tabular}{cc}
	\mbox{\hspace{-0.15in}}
        \epsfig{file=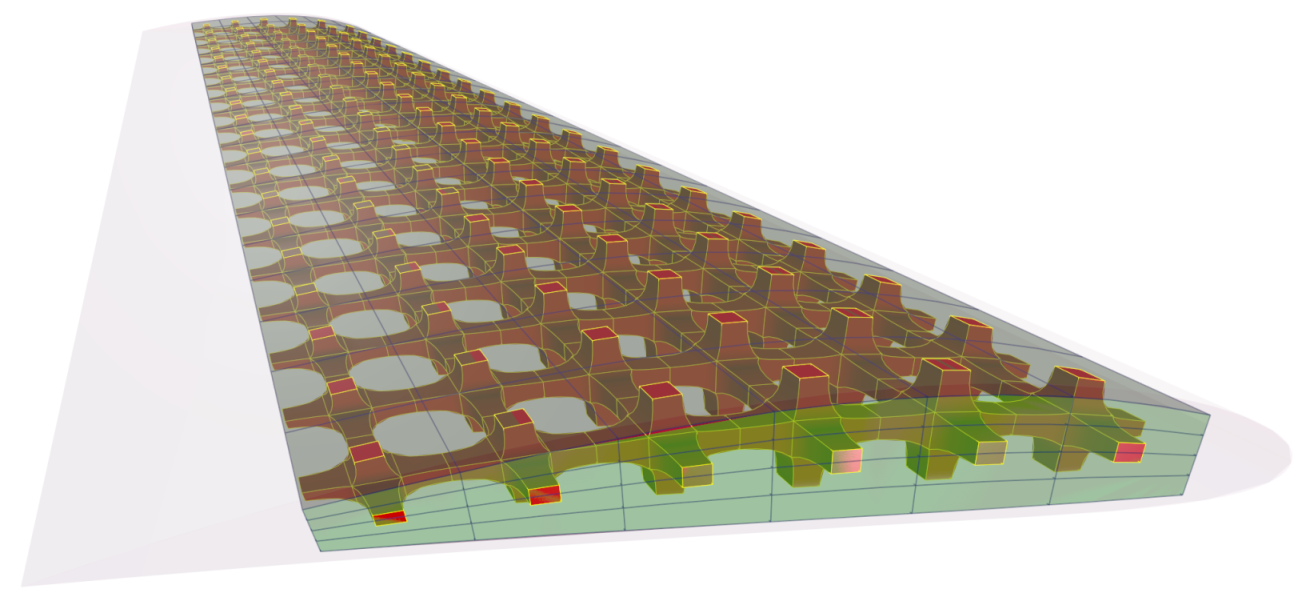,height=1.6in}
        \epsfig{file=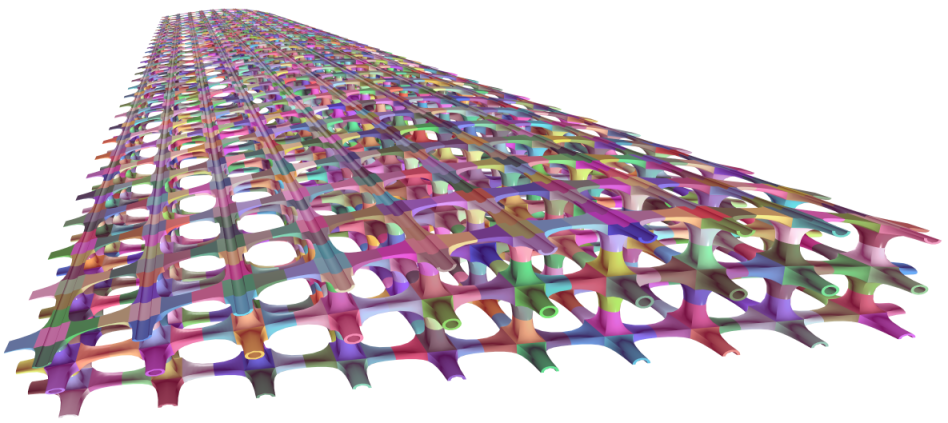,height=1.6in}
    \end{tabular}
    \begin{picture}(0,10)
        \put( -100, 95){(a)}
        \put(  185,100){(b)}
    \end{picture}
    \end{center}
    \vspace{-0.25in}
    \caption{One can model the interior of a wing, shown transparently in (a),
	     as a trivariate (also shown transparently in (a)),
	     in order to adaptively tile the trivariate
	     interior with micro-structures.
	     In (b), a more detailed structure, employing hollowed
	     tiles is presented.}
\label{fig-wing}
\end{figure*}

The use of additive manufacturing in building porous, lightweight
micro-structures has the potential for fundamentally changing wing
structures and therefore wing-structural design. Because the majority
of internal space in the wing remains open, routing wiring and piping
does not degrade the structural integrity. Fuel can be stored while
maintaining the structure by simply cordoning and walling off parts of
the micro-structural scaffolding and using the segregated region to
hold fuel. The wing's skin can be an integral part of manufacturing
(and not separately applied) by using tiles like those seen in
Figure~\ref{fig-parametric-tile-geom}. {\it Stiffeners} are not
required because the unsupported expanse of skin is now much smaller
(also the skin itself, as part of the tile, can be locally
thickened). The individual tiles can be hollowed to further reduce the
weight. See Figure~\ref{fig-wing}.

\begin{figure*}
    \begin{center}
    \begin{tabular}{cc}
       	\mbox{\hspace{-0.15in}}
        \epsfig{file=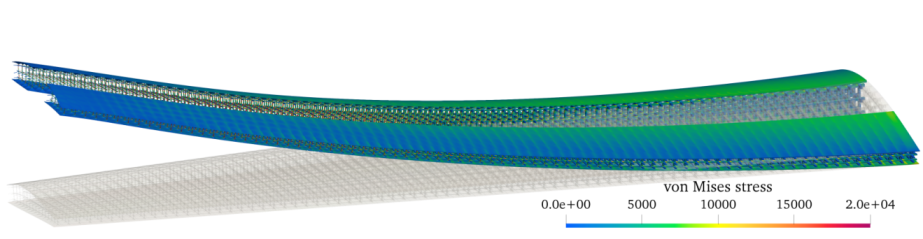,width=5.25in}\\
        \epsfig{file=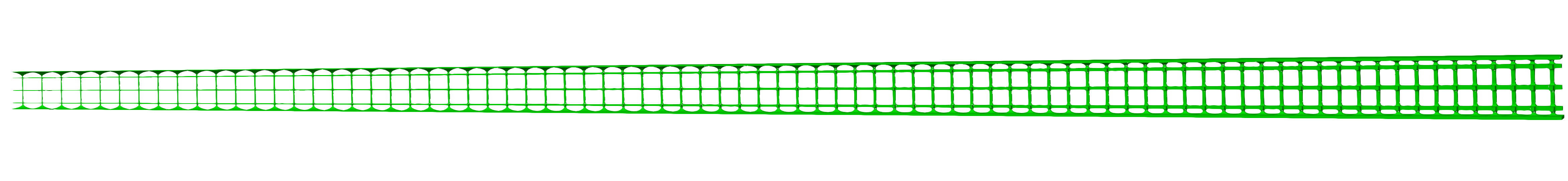,width=5.25in}\\
        \epsfig{file=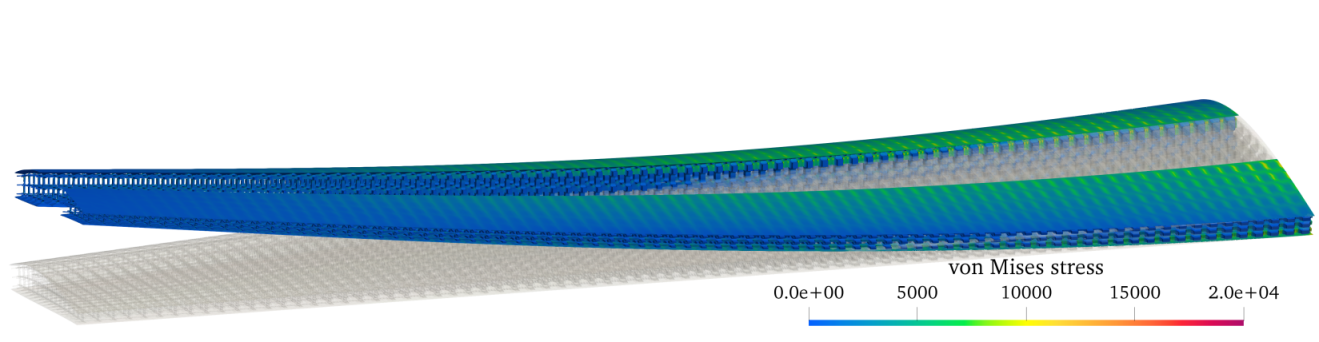,width=5.25in}\\
        \epsfig{file=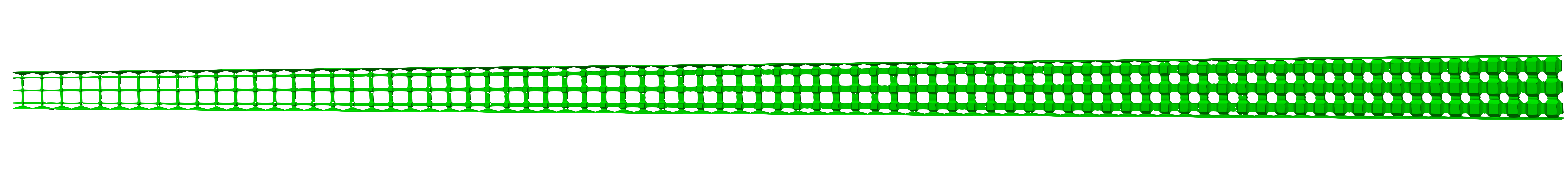,width=5.25in}\\
        \epsfig{file=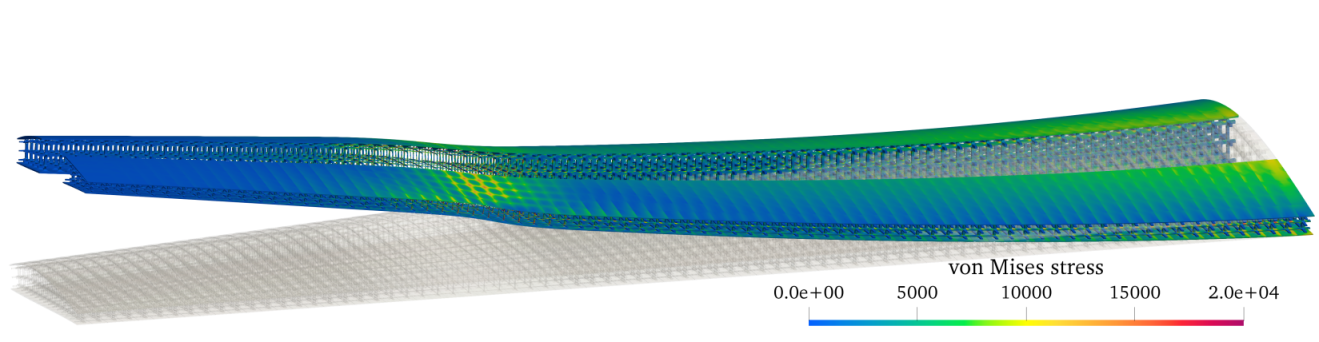,width=5.25in}\\
        \epsfig{file=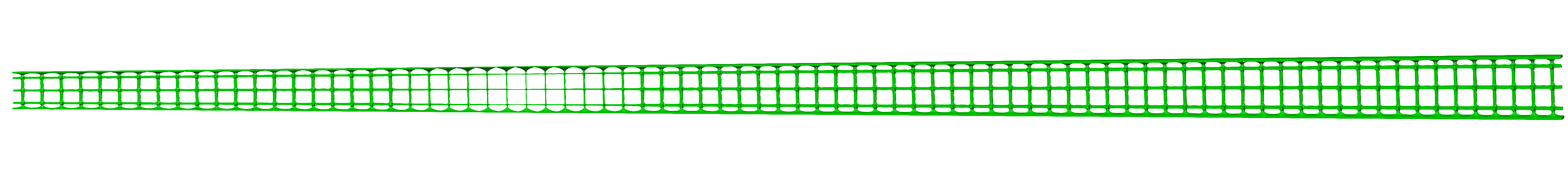,width=5.25in}\\
        \epsfig{file=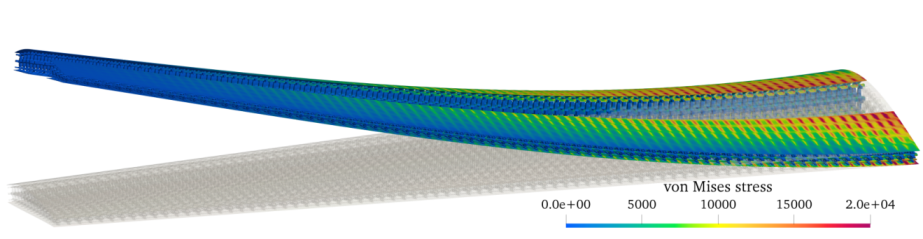,width=5.25in}\\
        \epsfig{file=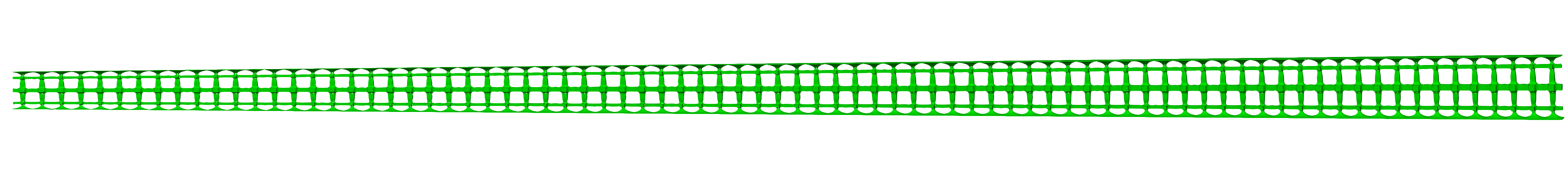,width=5.25in}
    \end{tabular}
    \begin{picture}(0,10)
        \put( -200, 190){(a)}
        \put( -200, 40){(b)}
        \put( -200, -110){(c)}
        \put( -200, -258){(d)}
    \end{picture}
    \end{center}
    \vspace{-0.25in}
    \caption{Structural analyses of different porous wing configurations, fixed at the root. For each case,
		a magnified elastic deformation is presented (under a homogeneous lift applied to the lower wing surface), colored with von Mises stress,
		together with a translucent representation of the non-deformed configuration.
		In order to show the wing interior, a longitudinal portion of the wing has been removed from the images.
		In addition, below each analysis a longitudinal section of the wing is included.
		The four different cases considered are:
		a) wing design with gradual change of tile thicknesses from the root (thick) to the tip (thin);
		b) same configuration as a), but using thicker tiles in the interior of the wing, and thinner near the skin;
		c) wing with constant thicknesses, except at a section situated at $2/3$ from the root;
		d) same as b), but without a gradual thickness reduction from the root to the tip.}
\label{fig-wing-results}
\end{figure*}
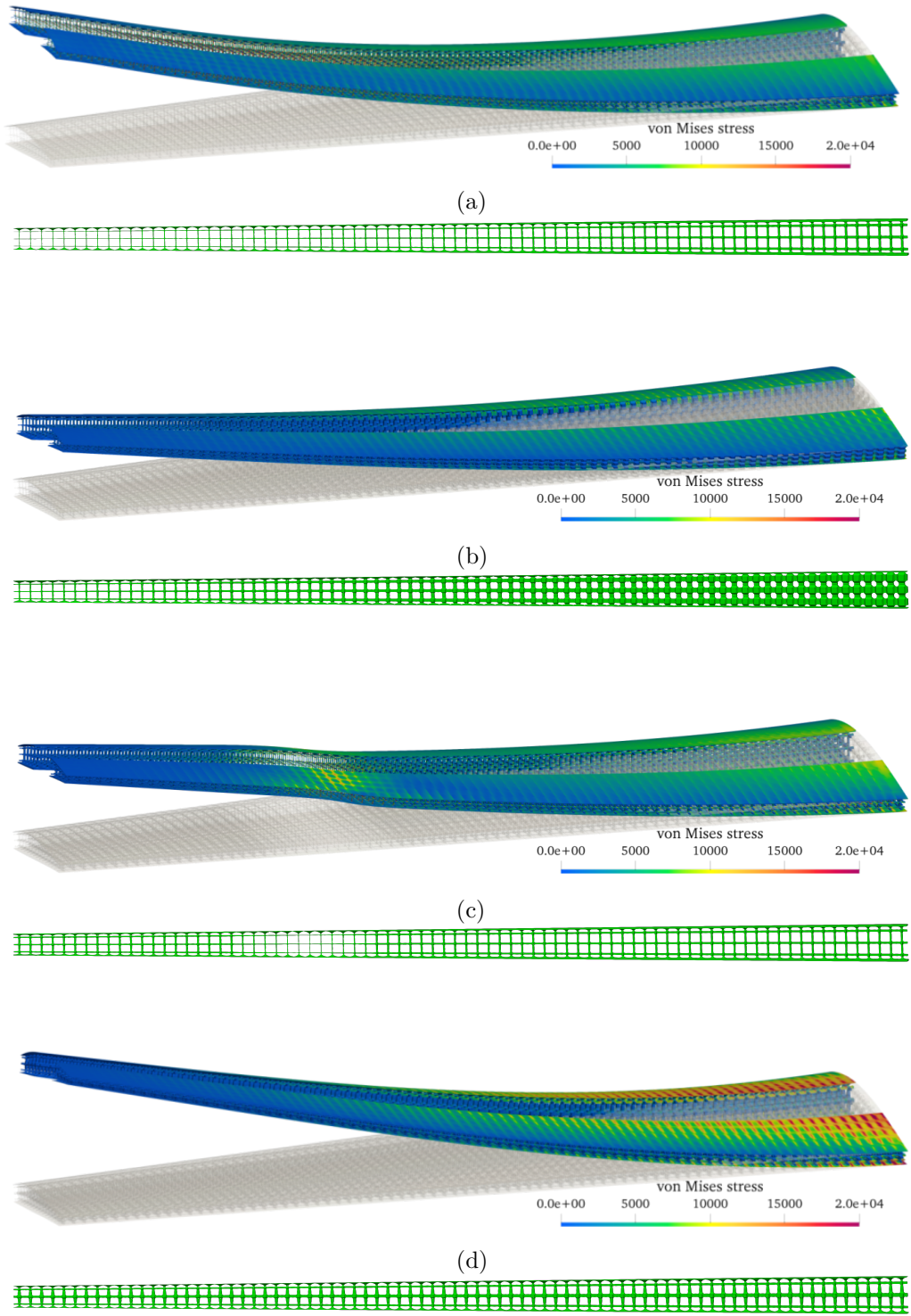

With the aim of exploring the basic structural response of porous wing
designs, four different configurations were analyzed in this section
(see results in Figure~\ref{fig-wing-results}).

All the cases considered present the same external wing profile and
number of tiles ($15\times80\times3=3600$), with a configuration
similar to those shown in Figure~\ref{fig-parametric-tile-geom}.

The same type of simulation was carried out for all designs: linear
elasticity analysis using the same trivariate parametrizations for the
description of the geometry and for the discretization of the
solution, following an isogeometric analysis framework (see
\cite{Cottrell2009} and references therein).

Each wing configuration presents a total of 51600 tricubic \Bezier{}
trivariates, thus, each model has 7 million unknowns: the elastic
displacements of each control point in the three Cartesian directions.
Contiguous tiles are conforming (the parametrizations of their common
faces coincide), therefore, the solution's $C^0$ continuity is imposed
by enforcing the equality of the unknown variables associated to
coincident control points\footnote{By enforcing the discrete elastic
displacement solution to be $C^0$ continuous at the interface between
connected trivariates, the number of unknowns is reduced to 7 million
from almost 10 ($51600\times64\times3$).}.  In general and much like
continuity preservation between surface patches, as long as one
preserves the continuity between adjacent tiles in the domain of the
deformation function (and the deformation function is sufficiently
continuous), the result will preserve continuity.  Finally, in the
case of non-conforming tiles, the solution's continuity could be
achieved in a weak way by means of mortaring techniques
\cite{Brivadis15}.  Using a multiple-core machine (70 threads were
used), the computational time was under 40 minutes in all cases.  The
computations were performed using the isogeometric analysis library
igatools~\cite{Igatools2015}.

In the structural analyses performed, the wing is fixed at the root,
the wider transversal section that connects the wing with the
aircraft's fuselage (the right extremity of the wing in
Figure~\ref{fig-wing-results}).  We considered a linear elastic
material with a Young modulus to Poisson ratio ratio $E/\nu=1/0.3$
(typical of steel-based materials).  Additionally, in a simplified
setting, a uniformly distributed pressure (lift), caused by the
aerodynamic loads, was applied on the bottom wing surface.  Under
these conditions, the wing bends vertically behaving roughly as a
cantilever beam: the higher stress concentration appears at the root,
whereas the largest wing deflection occurs at the tip.  However, as
shown in the obtained results, the reduction of the tile's thicknesses
can also induce stress concentrations, as it can be seen at thinner
section of the configuration c; or in the case d, in which the
presence of thinner tiles in the wing skin, near the root, cause an
increase of the stress level, with respect to other designs.

On the other hand, the oscillatory behavior of the stress
distributions at the wing skin follows the pattern defined by the
tiles distribution. In fact, the amplitude of these oscillations is
more accentuated in the case d, that presents a thinner external skin
layer.

As a summary, Table \ref{tab-wing} gathers the maximum wing
deflection, compared to the total volume of wing material (that will
determine the wing self-weight, of crucial importance for an aircraft
structure), for the four designs studied.

\begin{table}
\centering
\begin{tabular}{c|c|c}
Case & Relative volume  & Relative deflection \\\hline
a & $1.00$ & $1.76$ \\
b & $2.50$ & $1.00$ \\
c & $1.09$ & $1.31$ \\
d & $1.29$ & $2.06$ 
\end{tabular}
\caption{Relative wing material volume (the smaller, the better) respect to the lighter design (case a);
and relative wing maximum deflection (smaller is better) respect to most flexible design (case b).}
\label{tab-wing}
\end{table}

Looking at the table results, the heaviest design (case b) is also the
stiffest, as expected.  However, as it can be seen for cases a and c,
by optimizing the distribution of the tile's thicknesses, it could be
possible to achieve much lighter designs without significantly
reducing the wing's overall stiffness.\footnote{In a very simplified
setting, considering a solid wing design whose span and chord have a
fixed length, the maximum wing deflection can be estimated to be
proportional to the cube of its thickness (i.e., to the cube of its
volume).}

Further design improvements would benefit from the use of
multi-objective shape optimization techniques, in which the maximum
stress is minimized, while trying to keep a low total volume.  The
optimizer could operate globally, or over some neighborhood or even on
individual tiles, as long as the (geometric as well as material)
continuity between adjacent tiles is properly ensured.  Indeed, the
use of graded materials (see, e.g.,
\cite{stankovic_generalized_2015,Massarwi18}) potentially offers a
wide range of possibilities in this endeavor.

\subsection{Local Heating of Extruders}
\label{subsec-examples-heating-extruders}

Plastic profile extrusion is a manufacturing process particularly
suited for continuous profiles. These include pipes and floor
skirtings, but also more complex geometries, such as window panes. An
extrusion line consists of three important parts: (1) an extruder,
responsible for melting, mixing, and transporting the raw plastic, (2)
the extrusion die, responsible for reshaping the melt to the
desired profile, and (3) the calibration, which fixes the profile
shape during solidification. With its high influence on the quality of
the final product especially in terms of shape accuracy, the extrusion
die is certainly the component that has been investigated in most
detail. Quality criteria for an extrusion die revolve around shape
accuracy of the final product: This is, in particular, influenced by the
velocity distribution at the outflow of the extrusion die, as well as
the viscoelastic stresses induced within the die. Possible design
measures to influence these criteria are the shape of the flow channel
within the die, and also the temperature distribution.

In recent years, the research around extrusion dies has been directed
towards numerical design methods; i.e., the automated design of the
extrusion die, given a certain product shape as input. Ettinger gives
an overview of work regarding shape optimization
\cite{Ettinger2013}. More recent research can be found in
\cite{rajkumar2017, zhang2017, Pauli2013}. We note that all
of this work is focused on shape optimization of the flow
channel. So far, when it comes to simulations, the
temperature control within the extrusion die has mostly been
considered ideal: In practice, this means that the process is treated
as isothermal. This assumption is justified by the heating systems
that are currently used: The extrusion die is simply wrapped with a
heating band \cite{Michaeli1991}.It is clear that
such an approach gives no local control over the temperature.

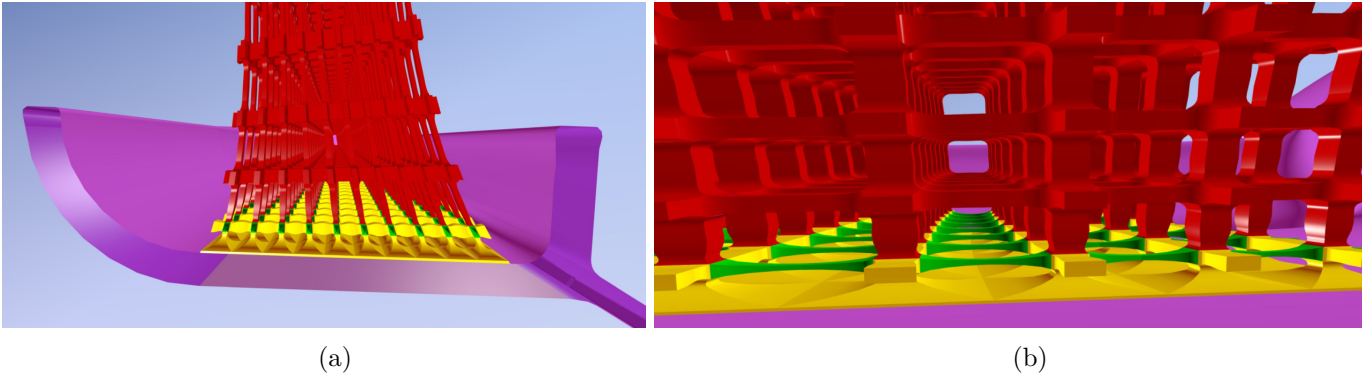
\begin{figure*}
    \begin{center}
    \begin{tabular}{c}
	\mbox{\hspace{-0.15in}}
        \epsfig{file=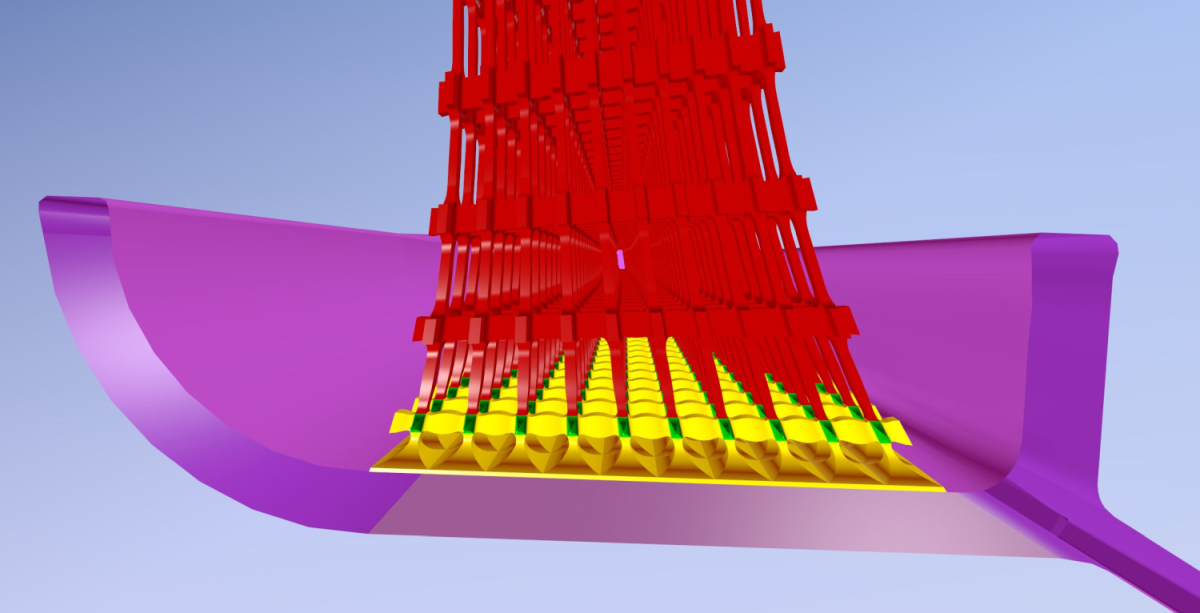,height=1.7in}
        \epsfig{file=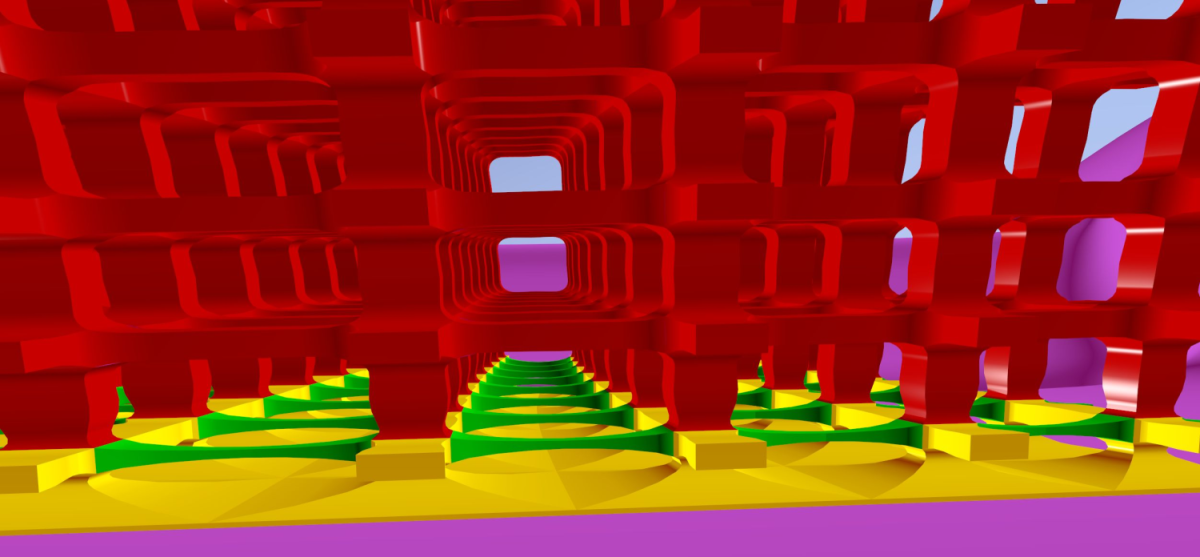,height=1.7in}
    \end{tabular}\\
    \begin{picture}(0,10)
        \put(-140, 0){(a)}
        \put( 120, 0){(b)}
    \end{picture}
    \end{center}
    \vspace{-0.25in}
    \caption{Local control over heating in an extruder (in magenta).
	     Red tiles are insulators (for both heat and electricity).
	     Yellow tiles are only thermally conductive, while green tiles
	     are electrically (and thermally) conductive.  Based on the
	     diameters of the cross sections of the green tiles, the
	     electrical resistance can be controlled, which locally
	     affects the amount of heat generated when electric current 
	     flows through the green tiles. (a) shows a view from
	     the outlet, while (b) shows a close-up side view.}
\label{fig-extruder-heating}
\end{figure*}

The micro-structuring approach proposed herein could - for the first
time - enable {\em local} temperature control; this would be achieved
without altering the overall mechanism for heating.  The extrusion die
can still be wrapped with the heating band, but the die would then be
micro-structured with material of graded heat conductivity. This would
allow for inhomogeneous temperature distributions within the flow
channel. As an example, confer with
Figure~\ref{fig-extruder-heating}. We see an extrusion die intended to
produce a floor skirting profile. A common problem arises when there
is a higher outflow velocity in the T-junction and a relatively low
outflow velocity at the tips of the profile, where wall-adhesion is
high due to the high surface area of the wall in this
regions. Mitigation of these effects based on flow-channel shape has
been investigated in \cite{Siegbert2013}.  A similar effect could be
achieved with locally reducing temperature in the T-junction, while,
at the same time, increasing temperature in the
tips. Figure~\ref{fig-extruder-heating} illustrates a possible
distribution of micro-structures that would result in the
aforementioned temperature distribution.

Furthermore, the use of micro-structuring opens up the realm of
developing systems using entirely new temperature control
approaches. For example, one could use variable inductive heating,
whereby the micro-structures are graded with respect to electrical
conductivity.

For this application, the microstructures will be parameterized in
terms of their geometry and their material properties. These can then
be modified locally with respect to the design objective of
homogeneous velocity distribution \cite{Elgeti2012}.  In order to keep
the number of optimization parameters manageable, expert knowledge
will be included into the parameterization. We envision the use of a
trust-region optimization algorithm like BOBYQA \cite{Powell2009}.

\section{Conclusions and Future Work}
\label{sec-conclusion}

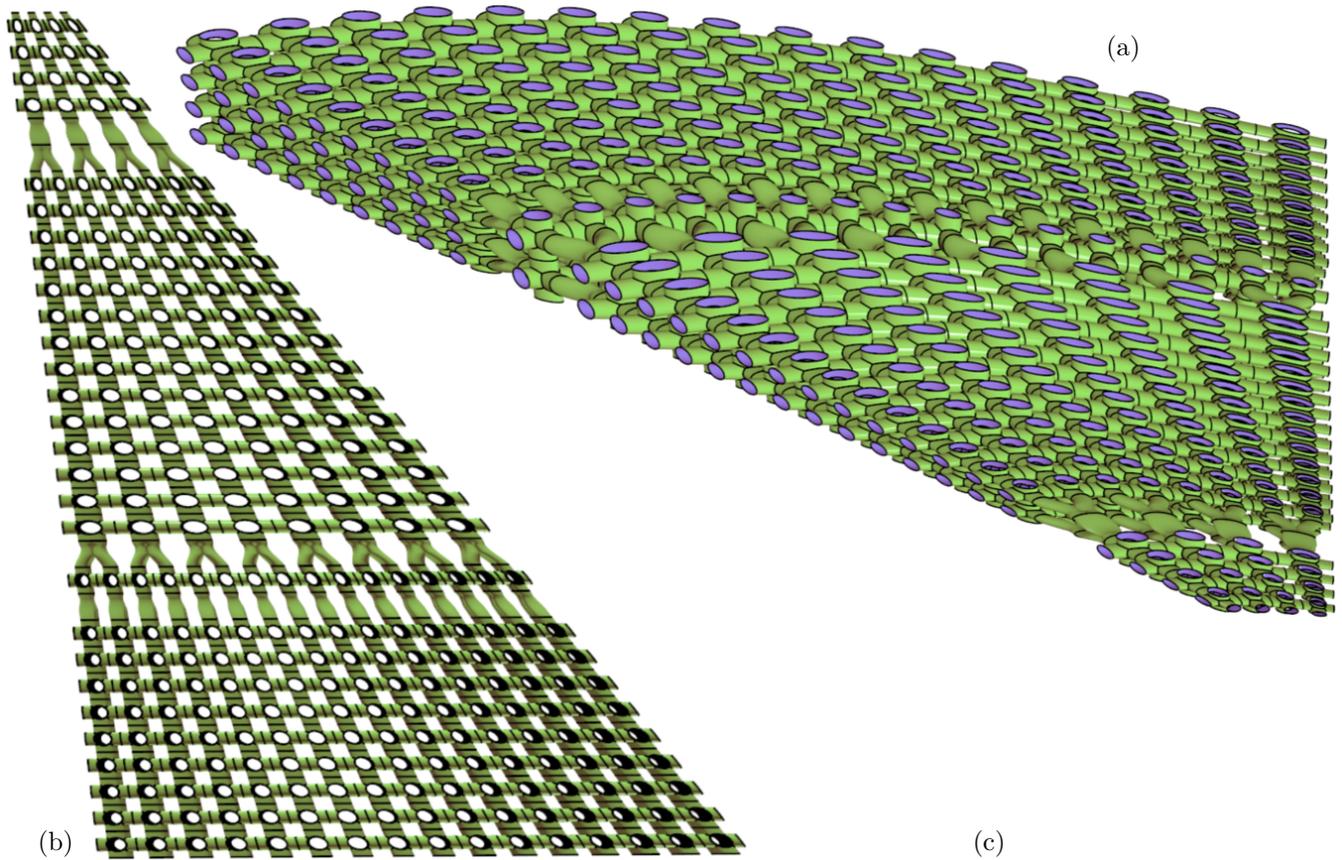
\begin{figure*}
    \begin{center}
    \begin{tabular}{c}
	\mbox{\hspace{-0.15in}}
        \epsfig{file=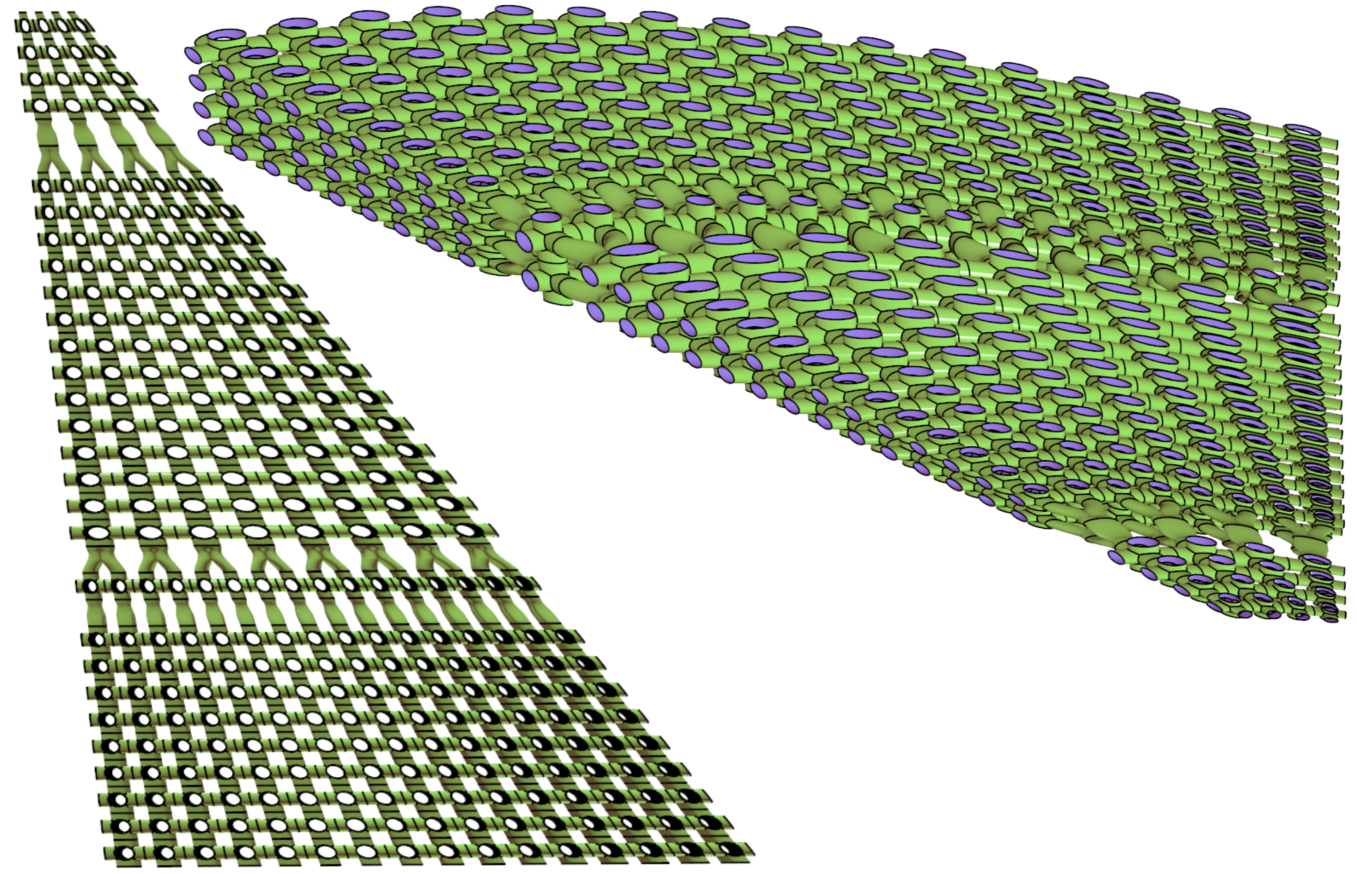,width=7in}
    \end{tabular}\\
    \begin{picture}(0,10)
        \put( 160, 320){(a)}
        \put(-240,  20){(b)}
        \put( 110,  20){(c)}
    \end{picture}
    \end{center}
    \vspace{-0.4in}
    \caption{Two views on a \Bspline{} surface micro-structure in the
	     shape of a delta wing with macro-tiles' surfaces that
	     employs bifurcations to change topology.  The surfaces
	     shown on the right view (a) are automatically merged by using
	     bifurcations in an effort to bound the minimal/maximal
	     tile size.  Starting from four rows of tiles, near the
	     root of the wing, it goes into two rows one third of the
	     way along the wing, and then into one row toward the tip
             of the wing.  Similar shrinkage in the number of tiles can
             also be observed from above (on the left in (b)).
	     (c) shows the a-priori defined parameteric
	     tiles with the different topologies.  The right-most tile
	     consists of trimmed surfaces whereas the other three tiles
	     of tensor product surfaces only.}
\label{fig-wing-delta}
\end{figure*}

\begin{figure}
    \begin{center}
    \begin{tabular}{c}
	\mbox{\hspace{-0.15in}}
        \epsfig{file=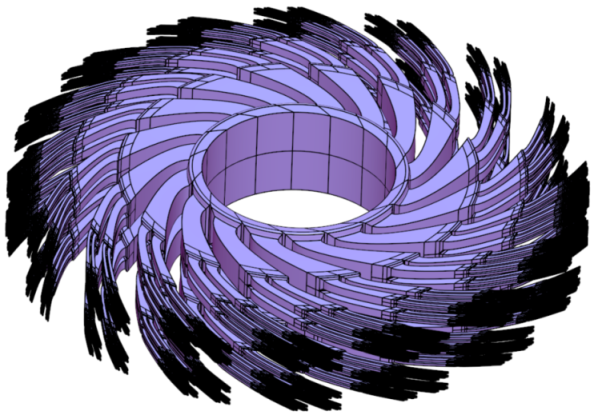,width=3.3in}
    \end{tabular}
    \end{center}
    \vspace{-0.25in}
    \caption{Another degree of freedom that can be exploited is the macro
	    shape of the micro-structure.  Here, the amount of
	    twisting of the ring is increased compared to
	    Figure~\protect\ref{fig-heat-sink}.}
\label{fig-heat-sink5}
\end{figure}

We have presented a design paradigm exploiting parametric tiling in
micro-structures.  Control over both the shape and the materials were
presented.  Further degrees of freedom to optimize include
\begin{itemize}
\item Control over the size of the micro-structure grid or recursive
      level or embedding nano-structures within micro-structures,
      etc.~\cite{Massarwi18}.
\item Control over the topology of the tiles, employing tiles with
      different topologies in different locations.  One such potential
      example is shown in Figure~\ref{fig-wing-delta} where
      bifurcations are employed.  This example employs
      tiles with (trimmed) surfaces and contiguous surfaces
      in the tiles that are not conforming.  For simpler analysis and
      optimization, it is better to design with conforming (preferably
      untrimmed) trivariate based tiles.
\item Control over the macro-shape or the deformation function.  One
      example for such an ability is shown in Figure~\ref{fig-heat-sink5}
      that synthesizes longer fingers to the heat sink.  Compare with
      Figure~\ref{fig-heat-sink}.
\end{itemize}

The example in Figure~\ref{fig-wing-delta} demonstrates the potential
in an effort to keep all tiles of uniform size.  Such a constraint is
typically application driven (i.e. minimal thickness walls in additive
manufacturing) and is not a limit of the presented process.  The
micro-tiles can be deformed arbitrarily.  Further, as many tiles as
desired can fit into one deformation macro-function.  It is limited
only by computer memory needs and computational costs.

Up to continuity requirments, the different tiles in the domain of the
deformation macro-function are independent and can be arbitrarily
different (and even random).  Each tile can present a different
topology, geometry or material properties.  How this generality will
be fully exploited in design is yet to be seen.

There has been a great deal of recent interest in the use of
Topological Optimization to generate interesting (and sometimes
unintuitive) structural designs. This focus is partially due to the
commonalities with additive manufacturing that both use the same
volumetric underpinning (voxels). This means that a structural design
can be simply printed without translation. But like any engineering
tool there are limitations, which in this case include: deep optimal
designs (which can be fragile), difficulties when surface smoothness
is critical, compatibility with contemporary CAD systems, and dealing
with designs where (structural) analysis is only one of many
disciplines in play.

It is the last point above (designs for multi-physics devices) when
micro-structures provide a general viable alternative to voxels and
therefore the possibility to design in multidisciplinary
settings. Examples of this can be seen throughout
Section~\ref{sec-examples} where the outer/macro shape need not be
rectilinear and the micro-shapes have few limitations. The difficulty
in this design setting is that the number of parameters that drive the
design through optimization can be quite large (and some
continuous). This can be effectively handled in a gradient-based
optimization manner where the parametric derivatives for the entire
problem are available. Generating these derivatives can efficiently be
accomplished through the use of tightly coupled physics solvers that
include their Adjoint so that the full Jacobian of the coupled problem
can be made available. Then the chain-rule can be applied to couple
the Jacobian to the parametric derivatives produced from
differentiating the geometry construction.

\section*{Acknowledgments}
\label{sec-ack}

This research was supported in part by the ISRAEL SCIENCE FOUNDATION
(grant No. 597/18) and in part with funding from the Defense
Advanced Research Projects Agency (DARPA), under contract
HR0011-17-2-0028. The views, opinions and/or findings expressed are
those of the author and should not be interpreted as representing the
official views or policies of the Department of Defense or the
U.S. Government.
Pablo Antolin and Annalisa Buffa gratefully acknowledge the support
of the European Research Council, through the ERC AdG n. 694515 - CHANGE. 

\section*{References}

\bibliographystyle{acm}

\bibliography{MicroStructDesign}

\end{document}